\documentclass[aps,prd,onecolumn,groupedaddress,showpacs,nofootinbib,amssymb]{revtex4}
\usepackage[dvips]{graphicx}
\usepackage{amssymb}
\usepackage{amsmath}
\usepackage{epstopdf}
\usepackage{epsfig}
\usepackage{graphicx,,color}
\usepackage{amsfonts}
\usepackage{bm}
\usepackage{cancel}
\usepackage{comment}
\allowdisplaybreaks[4]

\begin{document}

\tolerance=5000

%\documentclass[10pt,a4paper]{article}
%\usepackage[dvips]{color}
%\usepackage{epsfig}
%\usepackage{amsmath}
%\usepackage{graphicx}
%\usepackage{authblk}
%\usepackage{amssymb,amsmath}
%\usepackage{amsmath}
%\usepackage[belowskip=-10pt,aboveskip=0pt]{caption}
%\setlength{\intextsep}{10pt plus 5pt minus 2pt}

%\def\Box{\hbox{$\rlap{$\sqcup$}\sqcap$}}
%\textwidth=165 mm \textheight=220 mm \oddsidemargin=0 mm
%\parindent=10 mm
%\newcommand{\tp}{\tilde{\phi}}
%\begin{document}

\title{Cosmological singularities in interacting dark energy models with an $\omega(q)$ parametrization}

%\title{Interacting dark energy models with an $\omega(q)$ parametrization in light of Gaussian processes and H(z) data}

\author{
Emilio Elizalde$^{1,2,3}$\thanks{E-mail: elizalde@ieec.uab.es},
Martiros Khurshudyan$^{2,4,5,6}$\thanks{Email: khurshudyan@yandex.ru, khurshudyan@tusur.ru},
Shin'ichi Nojiri$^{3,7,8}$\thanks{E-mail: nojiri@gravity.phys.nagoya-u.ac.jp}}

\affiliation{
$^1$ Consejo Superior de Investigaciones Cient\'{\i}ficas, ICE/CSIC-IEEC,
Campus UAB, Carrer de Can Magrans s/n, 08193 Bellaterra (Barcelona) Spain \\
$^{2}$ International Laboratory for Theoretical Cosmology, Tomsk State University of Control Systems
and Radioelectronics (TUSUR), 634050 Tomsk, Russia \\
$^{3}$ Kobayashi-Maskawa Institute for the Origin of Particles and the Universe,
Nagoya University, Nagoya 464-8602, Japan \\
$^{4}$ Research Division,Tomsk State Pedagogical University, 634061 Tomsk, Russia \\
$^{5}$ CAS Key Laboratory for Research in Galaxies and Cosmology, Department of Astronomy, University of Science and Technology of China, Hefei 230026, China \\
$^{6}$ School of Astronomy and Space Science, University of Science and Technology of China, Hefei 230026, China \\
$^7$ Department of Physics, Nagoya University, Nagoya 464-8602, Japan \\
$^8$ KEK Theory Center, High Energy Accelerator Research Organization (KEK), Oho 1-1,
Tsukuba, Ibaraki 305-0801, Japan
}

\begin{abstract}

Future singularities arising in a family of models for the expanding Universe, characterized
by sharing a convenient parametrization of the energy budget in terms of the deceleration parameter,
are classified. Finite-time future singularities are known to appear in many cosmological scenarios, in particular, in the presence of viscosity or non-gravitational interactions, the last being known to be able to suppress or just change in some cases the type of the cosmological singularity. Here, a family of models with a parametrization of the energy budget in terms of the deceleration parameter are studied in the light of Gaussian processes using reconstructed data from $40$-value $H(z)$ datasets. Eventually, the form of the possible non-gravitational interaction between dark energy and dark matter is constructed from these smoothed $H(z)$ data. Using phase space analysis, it is shown that a non-interacting model with dark energy $\omega_\mathrm{de} = \omega_{0} + \omega_{1}q$ ($q$ being the deceleration parameter) may evolve, after starting from a matter dominated unstable state, into a de Sitter Universe (the solution being in fact a stable node). Moreover, for a model with interaction term $Q = 3 H b \rho_\mathrm{dm}$ ($b$ is a parameter and $H$ the Hubble constant) three stable critical points are obtained, what may have important astrophysical implications.  In addition, part of the paper is devoted to a general discussion of the finite-time future singularities obtained from direct numerical integration of the field equations, since they appear in many cosmological scenarios and could be useful for future extended studies of the models here introduced. Numerical solutions for the new models, produce finite-time future singularities of Type I or Type III, or an $\omega$-singularity, provided general relativity describes the background dynamics.
\end{abstract}


\maketitle

\section{Introduction}

Interacting dark energy cosmological models are attracting growing attention in recent times, owing to the fact that a consistent framework has been already developed to fit the observational data with increasing precision. There is also a trend to try to avoid introducing any kind of non-gravitational interaction, in particular, because these contributions do not easily follow from fundamental theories. However, it turns out  that in models without non-gravitational interactions it is usually very difficult to solve a number of key theoretical and cosmological problems in a convincing way. On the other hand, there is increasing evidence indicating that some kind of non-gravitational interaction between dark energy and dark matter might be imprinted in recently obtained observational data. Commonly, interacting dynamical dark energy models have been introduced in cosmology in order  to solve problems that the $\Lambda$CDM standard model has to face, when confronted with observations. In present day cosmology, dark fluids are often used as the basic objects and, in some sense, they successfully replace dynamical dark energy models. They can also emerge in modified theories of general relativity. One of the main aims of some important, ongoing research in cosmology is to find a realistic way to correctly parameterize the equation of state  (EoS)  of this dark fluid (also termed, the dark side of the Universe).

In the recent literature, increasing interest is revealed towards the study of interacting dark energy models and dark fluids,
see~\cite{Xia:2016vnp,Zhang:2013zyn,Yang:2017iew,Marsh:2016ynw,Sadjadi:2006qp,Hakobyan:2013hca,Sadeghi:2015nda,Khurshudyan:2013oba,Khurshudyan:2015mva,Khurshudyan:2015mpa,Khurshudyan:2016uql,Khurshudyan:2016dki,Khurshudyan:2016ziq,Khurshudyan:2016epw,Khurshudyan:2016atd,Khurshudyan:2016xst,Khurshudyan:2016zse,Khurshudyan:2017kmf,Khurshudyan:2017qtd,Brevik:2017msy,Elizalde:2017mrn,Nojiri:2005pu,Nojiri:2017opc,Elizalde:2016tmn,Brevik:2016kwq,Odintsov:2017icc,Bamba:2015uma,Nojiri:2017ncd,Kremer:2003vs,Brevik:2017juz,Capozziello:2005pa,Capozziello:2004ej,Gorini:2004by,Zhang:2004gc,Khurshudyan:2014fna,Copeland:2006tn,Cai:2009zp,Guberina:2006fy,Ludwick:2015dba,Capozziello:2005tf,Jhingan:2008ym,Khurshudyan:2017rrw}~(to mention only a few papers).
A particular example for the justification of the possible existence of a non-gravitational interaction between dark energy and dark matter can be found in Ref.~\cite{Abdalla:2014cla}. It is concluded there that the result reported by the BOSS experiment~\cite{Delubac:2014dt} for the Hubble parameter at $z=2.34$ is a direct evidence for the existence of a non-gravitational coupling between dark energy and dark matter. It should be mentioned that in Ref.~\cite{Abdalla:2014cla} a specific form of dark energy has been taken into account, which motivated the authors of Ref.~\cite{Khurshudyan:2017rrw} to seek other combinations of dark energy and dark matter, able to explain the results of the BOSS experiment, too. In particular, it has been shown in Ref.~\cite{Khurshudyan:2017rrw}  that a varying polytropic dark energy model with a logarithmic non-gravitational interaction  can also explain the mentioned result and yield theoretical results in very good correspondence with the PLANCK 2015 survey. In our opinion, it would be interesting to study the BOSS experimental results either without involving interacting dark energy models, or by using viscosity. The consideration of viscosity is appealing, as discussed in several reports, which clearly prove the effectiveness of involving viscosity into the background dynamics, not only for the late Universe but also when considering the early Universe and cosmic inflation. Studies of this kind are very important, since the BOSS experiment is unique in its nature and allows the scanning of the Universe at higher redshifts. At first glance, it seems that in order to explain the value of the Hubble parameter for $z=2.34$ at cosmological scales, it is necessary to introduce a dissipative process under the form of a non-gravitational interaction~(that is why we thought that viscosity could also be applicable). On the other hand, in the above mentioned studies the authors considered also parameterized forms of interacting dark energy models. But, in general we need to involve in this subject a model-independent analysis, because analysis driven from a strong dependence on the interacting dark energy pattern simply may fall. Therefore, organizing a study that uses a model-independent technique and allows a reconstruction of the parameters directly from observational data can be extremely valuable. In this paper we will adopt the Gaussian processes (GP) for our purposes~(see for instance ~\cite{MSeikel:2012ms} and references therein for more details on the method).
Gaussian processes constitute a powerful tool allowing to reconstruct the behavior of a function (and its derivatives) directly from given data, and have been used intensively to study various cosmological models. Moreover, studies carried out in recent literature have shown, for instance, that this method could also be applied to reconstruct the behavior of the non-gravitational interaction between dark energy and dark matter~(among other results).
Model-independent GP techniques depend on the covariance function (kernel), which can be estimated from the observational data. Therefore, for instance, they are able to model the deceleration parameter at different redshift directly from observational data. On the other hand, with the field equations describing the background dynamics, we can reconstruct other parameters interesting for completing the study. Before describing in detail the models of this paper and the datasets used, we would like to come back to the discussion of an interesting topic that is actively studied in recent literature.

The same reasons that motivated the consideration of modified theories of gravity, and the introduction of non-gravitational interactions, do not allow to find a solution in the case at hand. Sometimes, it has  been reported that the existing tension between different data sets gives an statistically equivalent picture from different scenarios; therefore, it is hard to make a final conclusion. However, it may be not excluded that the ultimate reason could be somehow deeper than assumed. Therefore, it is necessary to study different processes within the suggested cosmological models, in order to better understand possible differences among them. On the other hand, due to the obvious interest towards the fate of our Universe in the future, the study and classification of timelike future singularities, first undertaken in Ref.~\cite{Nojiri:2005sx}, is an alternative way aiming at a better understanding of the models. Intuitively, it is clear that if one of the parameters blows up, then a singularity should be formed. According to Ref.~\cite{Nojiri:2005sx}, the following types of singularities appear (for a few seminal references, see \cite{Caldwell:1999ew,Caldwell:2003vq,Barrow:2004xh,Barrow:2004hk,BouhmadiLopez:2006fu}):
\begin{itemize}
\item Type I~(``The Big Rip Singularity''). If the singularity occurs at $t = t_{s}$, then the scale factor $a$, the effective energy density $\rho_\mathrm{eff}$, and the pressure $P_\mathrm{eff}$, diverge as $t \to t_{s}$; that is, $a \to \infty$, $\rho_\mathrm{eff} \to \infty$ and $|P_\mathrm{eff}| \to \infty$. This case yields incomplete null and time-like geodesics,
 \item Type II~(``The Sudden Singularity''). In this case, the scale factor $a$ and the total effective energy density $\rho_\mathrm{eff}$ are finite, but the effective pressure $P_\mathrm{eff}$ diverges as $t \to t_{s}$; that is, $a \to  a_{s}<\infty$, $\rho_\mathrm{eff} \to \rho_{s}<\infty$ and $|P_\mathrm{eff}| \to \infty$. In this case the geodesics are complete and observers do not necessarily crush~(weak singularity),
\item Type III~(``The Big Freeze Singularity''). In this case, only the scale factor is finite, and the effective pressure and effective density diverge at $t \to t_{s}$; that is, $a \to  a_{s}<\infty$, $\rho_\mathrm{eff} \to \infty$ and $|P_\mathrm{eff}| \to \infty$. These can be either weak or strong singularities, which are geodesically complete solutions.
\item Type IV~(``Generalized Sudden Singularity''). In this case, the scale factor, the effective pressure and the effective density are finite at $t \to t_{s}$. The Hubble rate and its first derivative are also finite, but the higher derivatives of the Hubble rate diverge at $t \to t_{s}$. In this case a weak singularity appears and geodesics are complete.
\end{itemize}
Increasing effort is being devoted towards the classification of singularities in all sort of cosmological models, where
$P_\mathrm{eff} = \sum P_\mathrm{i}$ and $\rho_\mathrm{eff} = \sum \rho_\mathrm{i}$ are the effective pressure and the energy density made by different fluids with $P_\mathrm{i}$ and $ \rho_\mathrm{i}$ existing in the Universe.  In particular, there is an intensive study of cosmic singularities in the case of modified gravity theories. Discussions of these studies on modified theories and on the classification of singularities in general can be found in~\cite{Nojiri:2017ncd,Nojiri:2006ri,Nojiri:2010wj,Elizalde:2010ep,Elizalde:2010jx,Cognola:2007zu,Briscese:2006xu,Elizalde:2004mq,Harko:2011kv,Nojiri:2006gh,Oikonomou:2016jjh,Astashenok:2013vza,Astashenok:2017dpo,Linder:2010py,Ferraro:2006jd,Bamba:2016wjm,Chen:2010va,Myrzakulov:2010tc,Dent:2011zz,Cai:2011tc,Sharif:2011bi,Odintsov:2016plw,Odintsov:2015gba,Odintsov:2015zza,Bamba:2008ut}~(and references therein). Also we would like to mention the recent paper~\cite{Beltran:2016jb} where the authors try to establish a general relation between interactions of a specific form introduced at the level of the continuity equations and future cosmological singularities. In particular, their study shows that every finite-time future cosmological singularity can be directly mapped into a singularity of the interaction term, which they have dubbed as Q-singularity. It should be mentioned that the authors have considered specific parameterizations of the Hubble expansion rate and the dark energy equation of state (EoS). Another interesting result presented by these authors is a novel type of singularity characterized by a divergence in the time-derivative of the EoS parameter for dark energy. Such singularity is expected not to be relevant for the background evolution, but it might signal the presence of divergences in the sound speed of the perturbations. It seems clear that a mapping should exist, as mentioned. In particular, having in mind the form of the field equations describing the background dynamics and the specific parametrization, it is a matter of calculus to obtain the mapping. Depending on the form of the assumed parametrization, the mapping may not be one to one. We think that this could be observed in the case of non-linear interactions. In summary, it seems important to extend this study and obtain a mapping not relying on the parametrization used, i.e., to establish a general, parametrization independent mathematical framework and device an algorithm to explore this mapping.

In this paper a new parametrization of the EoS parameter corresponding to the dark energy/fluid is introduced based on a model earlier suggested in~\cite{Khurshudyan:2014yoa}. In this paper we will adopt two ways to study the models. In the first case we will use GP to reconstruct the behavior of the Hubble parameter from the data and study the models. While, in the second case, we will numerically solve the field equations for given initial conditions and then discuss the results which cannot be obtained with a GP method.  In the last case, the results will be presented as a motivation for a future study of the corresponding models. In particular, after some analysis of the numerical solutions, which shows that the models can explain the accelerated cosmic expansion and also solve the cosmological coincidence problem, the classification of all possible timelike future singularities arising in the models will be undertaken based on the imposed initial conditions and constraints. As a result of these constraints the final fate of the present Universe will be determined, according to these models, by a Type I, Type III, or an $\omega$-singularity~(with $a \to \infty$, $\rho_\mathrm{eff} \to 0$ and $|P_\mathrm{eff}| \to 0$, with EoS parameter $\omega \to \infty$, for $t \to t_{s}$)~\cite{Dabrowski:2009kg,FernandezJambrina:2007sx}.

Before  analyzing the numerical solutions, we will use a model-independent data smoothing GP algorithm on a $30$-point sample deduced from the differential age method for $H(z)$ and, in addition, on a $10$-point sample obtained from the radial BAO method data~(see Table~\ref{tab:Table0}). Furthermore, directly from the reconstructed $H(z)$ we can reconstruct the behavior of the deceleration parameter, which in terms of $H$ and $H^{\prime}$ reads
\begin{equation}
q(z) = -1 + \frac{1+z}{H}H^{\prime},
\end{equation}
where $\prime$ means $d/dz$. In other words, we can reconstruct a model-independent behavior of the deceleration parameter, which will be used also to reconstruct the behavior of the EoS parametrization and of other important cosmological parameters.

\begin{table}[t]
  \centering
    \begin{tabular}{ |  l   l   l  |  l   l  l  | p{2cm} |}
    \hline
$z$ & $H(z)$ & $\sigma_{H}$ & $z$ & $H(z)$ & $\sigma_{H}$ \\
      \hline
$0.070$ & $69$ & $19.6$ & $0.4783$ & $80.9$ & $9$ \\

$0.090$ & $69$ & $12$ & $0.480$ & $97$ & $62$ \\

$0.120$ & $68.6$ & $26.2$ &  $0.593$ & $104$ & $13$  \\

$0.170$ & $83$ & $8$ & $0.680$ & $92$ & $8$  \\

$0.179$ & $75$ & $4$ &  $0.781$ & $105$ & $12$ \\

$0.199$ & $75$ & $5$ &  $0.875$ & $125$ & $17$ \\

$0.200$ & $72.9$ & $29.6$ &  $0.880$ & $90$ & $40$ \\

$0.270$ & $77$ & $14$ &  $0.900$ & $117$ & $23$ \\

$0.280$ & $88.8$ & $36.6$ &  $1.037$ & $154$ & $20$ \\

$0.352$ & $83$ & $14$ & $1.300$ & $168$ & $17$ \\

$0.3802$ & $83$ & $13.5$ &  $1.363$ & $160$ & $33.6$ \\

$0.400$ & $95$ & $17$ & $1.4307$ & $177$ & $18$ \\

$0.4004$ & $77$ & $10.2$ & $1.530$ & $140$ & $14$ \\

$0.4247$ & $87.1$ & $11.1$ & $1.750$ & $202$ & $40$ \\

$0.44497$ & $92.8$ & $12.9$ & $1.965$ & $186.5$ & $50.4$ \\

%$$ & $$ & $$ & $$ & $$ & $$\\
\hline
$0.24$ & $79.69$ & $2.65$ & $0.60$ & $87.9$ & $6.1$ \\
$0.35$ & $84.4$ & $7$ &  $0.73$ & $97.3$ & $7.0$ \\
$0.43$ & $86.45$ & $3.68$ &  $2.30$ & $224$ & $8$ \\
$0.44$ & $82.6$ & $7.8$ &  $2.34$ & $222$ & $7$ \\
$0.57$ & $92.4$ & $4.5$ &  $2.36$ & $226$ & $8$ \\
          \hline
    \end{tabular}
    \vspace{5mm}
\caption{$H(z)$ and its uncertainty $\sigma_{H}$, in units of km s$^{-1}$ Mpc$^{-1}$. The upper panel corresponds to the $30$-point sample deduced from the differential age method. The lower panel, to the $10$-point sample obtained from the radial BAO method. See~\cite{Zhang:2016zh} and references therein for the details.}
  \label{tab:Table0}
\end{table}

The covariance function (kernel) $k(z,\hat{z})$ which correlates the function $H(z)$ at different points is chosen to be a squared exponential
\begin{equation} \label{eq:kernel}
k(z,\hat{z}) = \sigma_{f}^{2} \exp \left[ -\frac{|z-\hat{z}|^{2}}{2l^{2}} \right],
\end{equation}
with two hyperparameters $l$ and $\sigma_{f}$ to be determined from the observational data.

The models of dynamical dark fluid considered in this paper stem from a reparametrization of the EoS parameter by the deceleration parameter $q$~(for details see \cite{Khurshudyan:2014yoa})
\begin{equation}\label{eq:MainF}
P_\mathrm{de} = \omega_{0} q \rho_\mathrm{de},
\end{equation}
where
\begin{equation}
q = -1 - \frac{\dot{H}}{H^{2}}
\end{equation}
is the deceleration parameter. It is easy to see that such parametrization corresponds to ordinary matter for $\omega_{0} > 0$, at the early Universe, and to dark energy at the late time Universe, owing to the change of the sign of the deceleration parameter $q$. On the other hand, it is easy to see, after doing some mathematics, that the $\omega_{0}$ parameter for a Universe dominated by the fluid, Eq.~(\ref{eq:MainF}), can be expressed in terms of the deceleration parameter  $q_0$ in the present Universe, namely
\begin{equation}
\omega_{0} = \frac{2}{3}-\frac{1}{3 q_0}.
\end{equation}

In analyzing the models of this paper, it is assumed that the metric of the background corresponds to the flat Friedmann-Lema\^{\i}tre-Robertson-Walker (FLRW) universe, with line element
\begin{equation}
ds^{2} = -dt^{2} + a(t)^{2}\sum{(dx^{i})^{2}},
\end{equation}
where $a(t)$ is the scale factor of the Universe, while $R = 6(2H^{2} + \dot{H})$ stands for the Ricci scalar, with $H = \dot{a}/a$ being the Hubble parameter. The non-gravitational interaction is understood as energy transfer from one component to another, when they are interpreted as ideal fluids forming an effective fluid satisfying the following energy conservation law
\begin{equation}
\dot{\rho}_\mathrm{eff} + 3 H (\rho_\mathrm{eff} + P_\mathrm{eff}) = 0,
\end{equation}
where $\rho_\mathrm{eff} = \rho_\mathrm{de} + \rho_\mathrm{dm}$ and $P_\mathrm{eff} = P_\mathrm{de}$, since dark matter is assumed to be pressure-less. In this case, the non-gravitational interaction should be understood in the following way
\begin{equation}
\dot{\rho}_\mathrm{de} + 3 H (\rho_\mathrm{de} + P_\mathrm{de}) = -Q,
\end{equation}
and
\begin{equation}
\dot{\rho}_\mathrm{dm} + 3 H \rho_\mathrm{dm} = Q,
\end{equation}
where $Q$ is an accepted notation for the non-gravitational interaction, $H$ is again the Hubble parameter and the dot represents time derivative. Here $\rho_\mathrm{de}$ and $\rho_\mathrm{dm}$ are the energy densities for dark energy and dark matter, respectively, while $P_\mathrm{de}$ is the pressure of the dark fluid/energy.

The  paper is organized as follows. In Sect.~\ref{sec:CC} dark fluid  models based on a generalized parametrization of the EoS parameter in terms of the deceleration parameter $q$, are investigated. Then, some specific aspects of the phase space analysis are considered in detail, on top of the results obtained from the GP reconstruction. Sect.~\ref{sec:TS} is devoted to an exhaustive discussion of the results obtained directly from numerical analysis of the models. In addition, the important issue of the impact of various forms for the non-gravitational interaction on the type of the singularities appearing in the corresponding models and on its possible time evolution is considered. Finally, the last section, Sect.~\ref{sec:D}, is devoted to a final discussion and conclusions.

\section{The models, phase space analysis and GP processes}\label{sec:CC}

Phase space analysis is a very powerful tool intensively used in modern cosmology; there is indeed a huge number of works devoted to this subject. It allows, for instance, to demonstrate~(among other things) that the solution of the cosmological coincidence problem is directly due a non-gravitational interaction. In order to construct the phase portraits and obtain the critical points~(stable attractors), we introduce the new variables~(with $8\pi G = 1$)
\begin{equation}
x = \frac{\rho_\mathrm{de}}{3H^{2}},
\end{equation}
and
\begin{equation}
y = \frac{P_\mathrm{de}}{3H^{2}},
\end{equation}
and follow the common line of research to determine the critical points and to ascertain whether they are attractors or not. Some details justifying the need for these new variables and how to find the critical points and determine which of them is a late time attractor, for the background dynamics given by general relativity, can be found in Ref.~\cite{Khurshudyan:2015mva}, and references therein.

\begin{figure}[h!]
\begin{center}$
\begin{array}{cccc}
\includegraphics[width=125 mm]{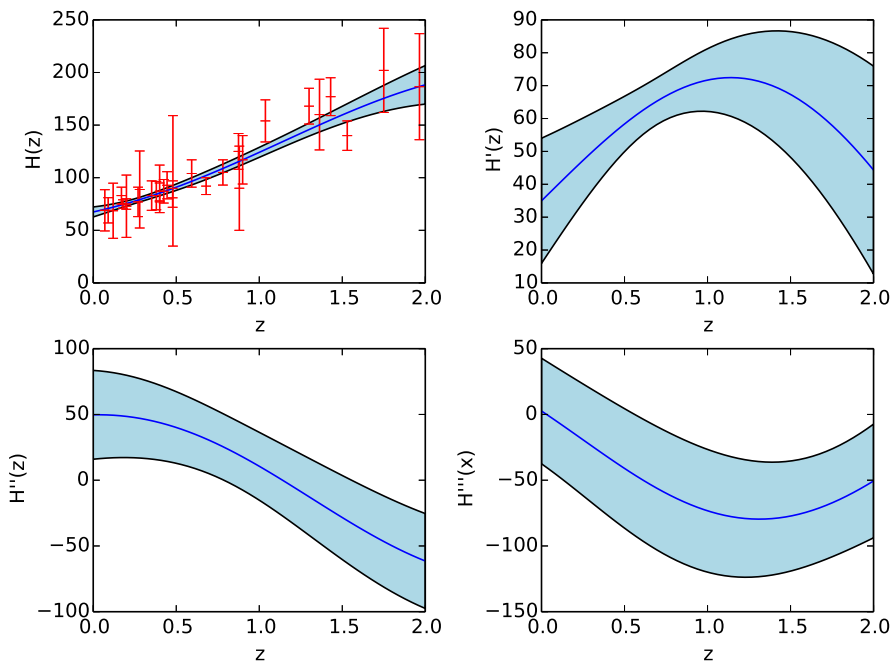}
\end{array}$
\end{center}
\caption{GP reconstruction of $H(z)$, $H^{\prime}(z)$, $H(z)^{\prime \prime}$, and $H(z)^{\prime \prime \prime}$ for the $30$-point sample deduced from the differential age method. The $^\prime$ means derivative with respect to the redshift $z$.}
 \label{fig:a}
\end{figure}

\begin{figure}[h!]
\begin{center}$
\begin{array}{cccc}
\includegraphics[width=125 mm]{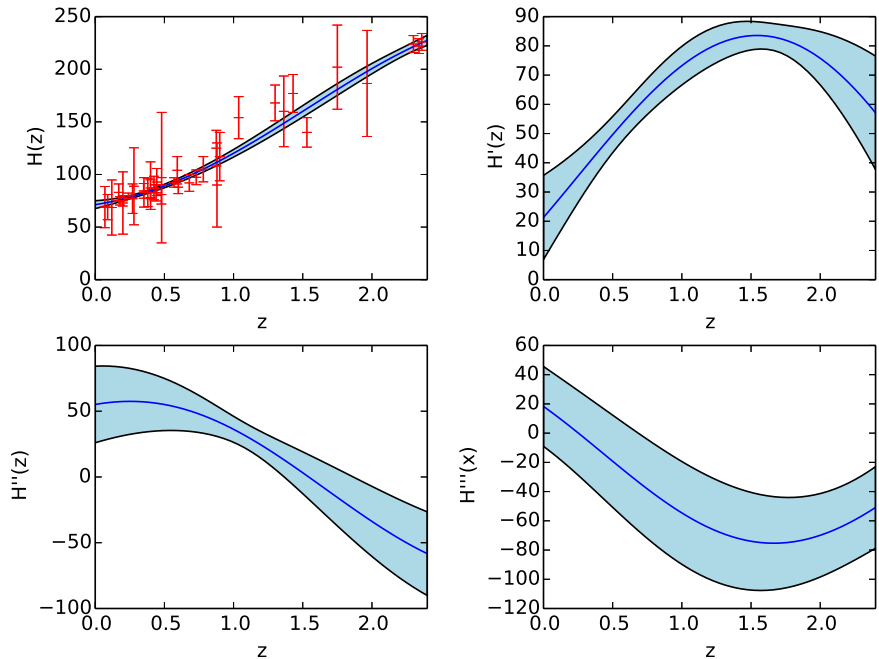}
\end{array}$
\end{center}
\caption{GP reconstruction of $H(z)$, $H^{\prime}(z)$, $H(z)^{\prime \prime}$, and $H(z)^{\prime \prime \prime}$ for the $30$-point sample deduced from the differential age method, with the additional $10$-point sample obtained from the radial BAO method. The $^\prime$ means derivative with respect to the redshift $z$.}
 \label{fig:b}
\end{figure}

On the other hand, GPs  have been also widely applied, in particular, to test the standard concordance model~\cite{Yahya:2014s} and the distance duality relation~\cite{Santos:2015s}. They have been used to reconstruct dark energy and cosmography~\cite{Holsclaw:2010th,Seikel:2012mcm,Shafieloo:2012as}. There is also interesting research aiming to determine the interaction between dark matter and energy~\cite{Yang:2015tzr}~(to mention a few applications only).  The choice for the kernel, Eq.~(\ref{eq:kernel}), seems logical due to the assumption on the distribution of observational data. On the other hand, in the above mentioned references there is also a detailed discussion on the reasons to consider other kernels, as well. 

In this paper we have used the public code GaPP (Gaussian Processes in Python) in order to study the models. In order to understand the weight of the suggested phenomenological parameterizations of the EoS, first we used only $30$-point samples of $H(z)$ deduced from the differential age method. Then, we added $10$-point samples obtained from the radial BAO method~(see Table~\ref{tab:Table0}). In the first case we have relatively good data up to $z=2$. On the other hand, in the second case we can extend the analysis up to $z=2.4$ improving also low-redshift data. In Fig.~\ref{fig:a} we present the reconstructed behavior of the Hubble parameter $H$ and its derivatives up to $z=2$, while Fig.~\ref{fig:b} represents the reconstruction up to $z=2.4$ using $30$ and $40$-point samples of $H(z)$ data, respectively. To be noted is that in both cases we avoid fixing the value of the Hubble parameter at $z=0$, i.e. the value of $H_\mathrm{0}$. A visual comparison of both plots already indicates that having additional high redshift data can significantly change and improve our understanding of the higher- and lower-redshift Universe. This is directly imprinted on the present day values of the Hubble and the deceleration parameters. In particular, the present day value of the Hubble parameter estimated from GP is  $71.286 \pm 3.743$ and $67.434 \pm 4.748$~($1 \sigma$ reconstruction level) for $40$ and $30$-point samples of $H(z)$ data, respectively. In our study besides the reconstructed behavior of the Hubble parameter $H$ we need to have the behavior of $H'$ and $H''$. The value of $H''$ is needed, in particular, in order to reconstruct  non-gravitational interactions.

\begin{figure}[h!]
\begin{center}$
\begin{array}{cccc}
\includegraphics[width=80 mm]{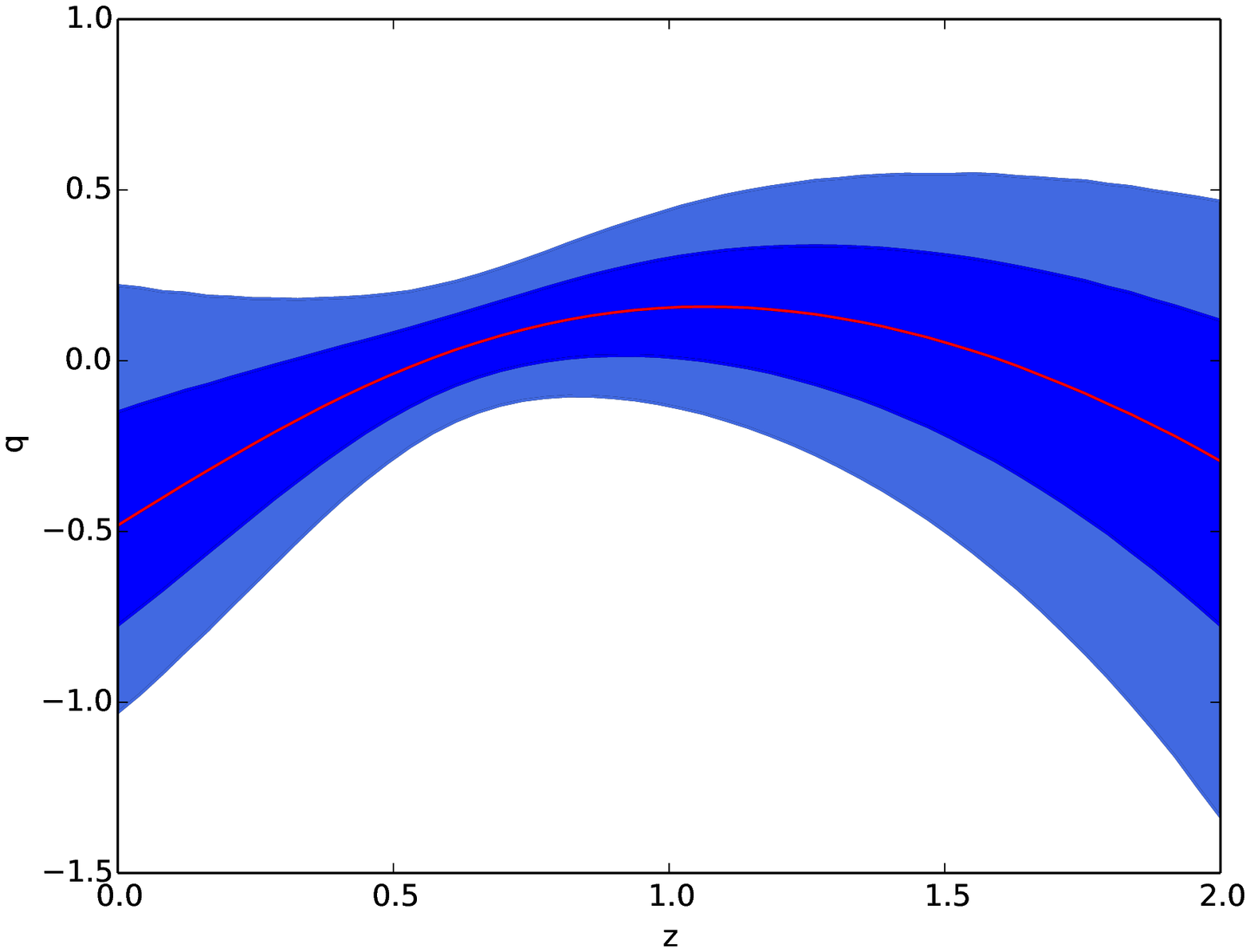}&
\includegraphics[width=80 mm]{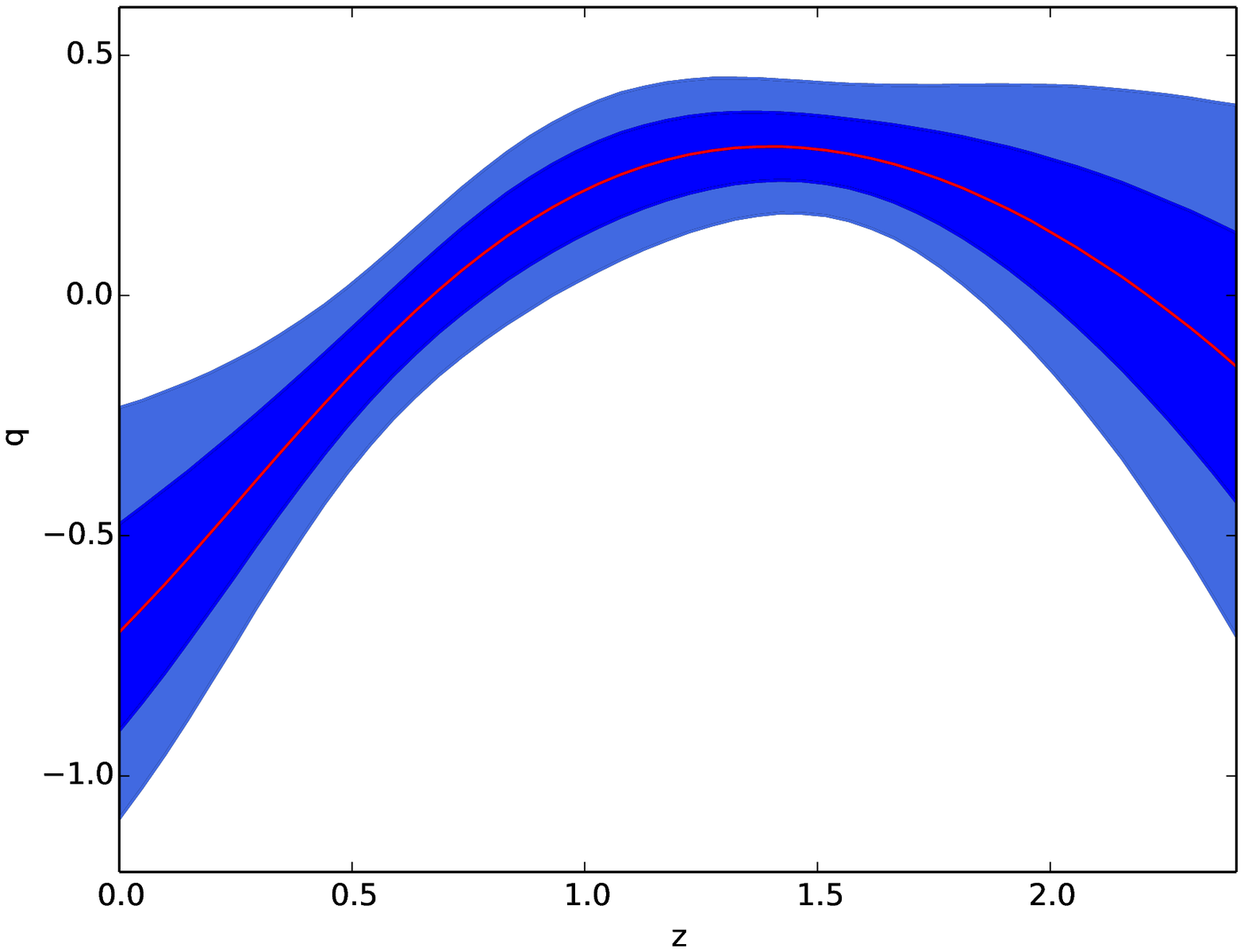}
\end{array}$
\end{center}
\caption{Reconstruction of the deceleration parameter $q(z)$ from the $H(z)$ data presented in Table~\ref{tab:Table0}. The left plot represents GP reconstruction from $30$-point $H(z)$ data, while the right plot presents the GP reconstruction from $40$-point $H(z)$ data. The solid line is the mean of the reconstruction and the shaded blue regions are the $68\%$ and $95\%$ C.L. bands of the reconstruction, respectively.}
 \label{fig:c}
\end{figure}

Complementarily, Fig.~\ref{fig:c} represents the reconstructed behavior of the deceleration parameter. Again, a direct and visual comparison of the behavior of the deceleration parameter presented in  Fig.~\ref{fig:c} demonstrates  the changes in the behavior of the deceleration parameter mainly due to higher redshift $H(z)$ data. It should be indicated clearly that the changes are due to a very high precision of the measurements~(small statistical error) of the $H(z)$ Hubble parameter at $z=2.30$, $z=2.34$ and $2.36$. Of course, it will be interesting to study how artificially increased statistical errors for the measurements at the mentioned redshifts will affect the conclusions of the study. This will be discussed elsewhere. From the two plots of Fig.~\ref{fig:c} we observe that the GP reconstruction of the deceleration parameter from $H(z)$ data gives rise to an accelerated expanding Universe for higher redshifts~($q<0$). Moreover, it is expected also to have a transition from this phase to the phase with $q > 0$, and for relatively lower redshift, again, we will have an accelerated expanding Universe. We also observe, that the present day value of the deceleration parameter is very sensitive and can be significantly changed depending on the considered data points.
In this paper we also will reconstruct the non-gravitational interaction between dark energy and dark matter in order to address the question of including a non-gravitational interaction to explain the value of the Hubble parameter at $z=2.34$, as reported by the BOSS experiment. Therefore, in the next two subsection we will concentrate our attention mainly on the results obtained from the reconstruction from the full $40$-point sample of $H(z)$ data. It is obvious that in this case the reconstructed non-gravitational interaction will answer the relevant question: do we need to include a non-gravitational interaction to explain the value of the Hubble parameter at $z=2.34$? In addition, in parallel we will discuss also important aspects arising from the $30$-point sample of $H(z)$ data.

\subsection{Model with $\omega_\mathrm{de} = \omega_{0} + \omega_{1}q$}\label{subsec:CC_1}

In this section we study the cosmology of a particular model of dark fluid in the presence of dark matter. The model of dark fluid is given by the following EoS parameter
\begin{equation}\label{eq:omegaDE}
\omega_\mathrm{de} = \omega_{0} + \omega_{1}q,
\end{equation}
where $\omega_{0}$ and $\omega_{1}$ are parameters of the model that are to be determined by using observational data, while $q$ is the deceleration parameter and $q<0$ corresponds to accelerated expansion. Moreover, for the Universe where the background dynamics is dominated by the barotropic dark fluid $P_\mathrm{de} = \omega_\mathrm{de} \rho_\mathrm{de}$, with EoS parameter given by Eq.~(\ref{eq:omegaDE}), the $\omega_{0}$ parameter can be easily expressed in terms of $\omega_{1}$ and the deceleration parameter  $q_0$ in the present Universe, in the following way
\begin{equation}
\omega_{0} = \frac{1}{3} \left (-1 + q_0 (2-3 \omega_{1}) \right ).
\end{equation}

On the other hand, it is easy to see that for this model the autonomous system
\begin{align}
y^{\prime} =& \frac{dy}{dN} = \frac{6 \omega_{0} (x+y)+3  \omega_{1} (3 y+1) (x+y)-6 y (y+1)}{3 x \omega_{1}-2} , \nonumber \\
x^{\prime} =& \frac{dx}{dN} = 3 (-1 + x) y,
\end{align}
where $N = \ln a$ is the folding time, has three critical points: $(x_{1},y_{1}) = (1,-1)$, $(x_{2},y_{2}) = (0,0)$, and
\begin{equation}
(x_{3},y_{3}) = \left (1,\frac{2 \omega_{0}+\omega_{1}}{2-3 \omega_{1}} \right ).
\end{equation}

Moreover, it comes out  that $(x_{2},y_{2}) = (0,0)$ is an unstable solution representing a matter dominated Universe, while the $\omega_{0} - \omega_{1}$ parameter space can be divided into several regions, according to  whether they are stabilized by  either the $(x_{1},y_{1})$ solution, or the $(x_{3},y_{3})$ solution. In other words, in the model considered the evolution should start from the $(x_{2},y_{2})$ state and end up either on $(x_{1},y_{1})$, or on $(x_{3},y_{3})$. For instance, with $\omega_{0},\omega_{1} \in [-10,10]$ both solutions are stable nodes for appropriate regions of the parameter space. It should be mentioned that the $(x_{1},y_{1})$ solution corresponds to a de Sitter Universe with $q = -1$ and $\omega_\mathrm{de} = \omega_\mathrm{eff} = -1$. On the other hand, the $(x_{3},y_{3})$ solution represents the state of the Universe with $\omega_\mathrm{de} = \omega_\mathrm{eff} = (2 \omega_{0}+\omega_{1})/(2-3 \omega_{1})$ and
\begin{equation}
q = \frac{3 \omega_{0}+1}{2-3 \omega_{1}}.
\end{equation}

\begin{figure}[h!]
\begin{center}$
\begin{array}{cccc}
\includegraphics[width=95 mm]{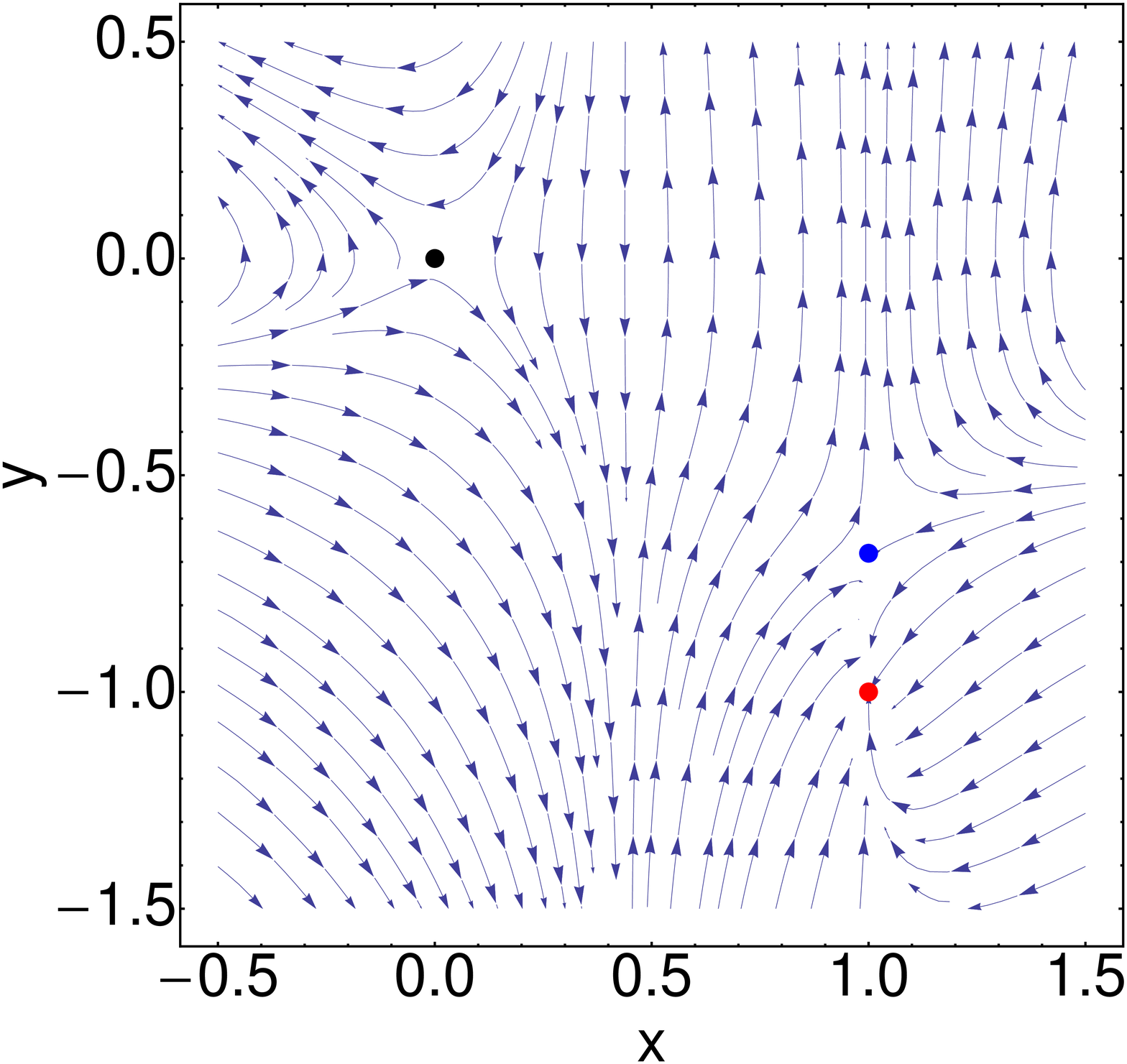}
\end{array}$
\end{center}
\caption{Phase space portrait for the cosmological model with a dark fluid described by Eq.~(\ref{eq:omegaDE}). The red dot represents the ($x_{1}$, $y_{1}$) solution, while the ($x_{2}$, $y_{2}$) and ($x_{3}$, $y_{3}$) solutions are represented by black and blue dots, respectively. We have set $\omega_{0} = 0.1$ and $\omega_{1} = 1.5$, and remember that $x\in [0,1]$.}
 \label{fig:0}
\end{figure}

\begin{table}[t]
  \centering
    \begin{tabular}{ | l | l | l | l  | l | l | l | p{2cm} |}
    \hline
C.P. & $x$ & $y$ & $q$ & $\omega_\mathrm{de}$ & $\omega_\mathrm{eff}$ & $\Omega_\mathrm{de}/\Omega_\mathrm{dm}$ & Type of stability \\
      \hline
C.P.1 & $\frac{1}{b+1}$ & $-1$ & $-1$ & $-b-1$ & $-1$ & $\frac{1}{b}$ & Stable Node or Focus\\
          \hline
C.P.2 & $\frac{2 (b+\omega_{0})+\omega_{1}}{3 b \omega_{1}+2 \omega_{0}+\omega_{1}}$ & $\frac{2 (b+\omega_{0})+\omega_{1}}{2-3 \omega_{1}}$ & $\frac{3 b+3 \omega_{0}+1}{2-3 \omega_{1}}$ & $\frac{3 b \omega_{1}+2\omega_{0}+\omega_{1}}{2-3 \omega_{1}}$ & $\frac{2 (b+\omega_{0})+\omega_{1}}{2-3\omega_{1}}$ & $\frac{2 (b+\omega_{0})+\omega_{1}}{b (3 \omega_{1}-2)}$ & Stable Node\\
          \hline
    \end{tabular}
    \vspace{5mm}
\caption{Two stable critical points (C.P.) corresponding to the interacting model with $Q = 3Hb\rho_\mathrm{de}$. The type of stability has been analyzed imposing the following prior constraints: $\omega_{0} \in [-1,1]$, $\omega_{1} \in [-1,1]$, and $b \geq 0$.}
  \label{tab:Table1}
\end{table}

However, according to recent observational data the last solution is less favorable since it describes an accelerated expanding Universe with $q < -1$. An example of the phase portrait for this model is presented in Fig.~\ref{fig:0} indicating that $(x_{2},y_{2})$ and ($x_{3}$, $y_{3}$) are saddle points, while ($x_{2}$, $y_{2}$) is a stable node. Table~\ref{tab:Table1} shows two critical points obtained for the cosmological model with $Q = 3H\rho_\mathrm{de}$ non-gravitational interaction between dark energy and dark matter. Also, the model has one unstable critical point similar to the one corresponding to the non-interacting model~($x=0$ and $y=0$). Therefore, starting the evolution from that state the Universe  will lead us  to the Universe as described either by the critical point C.P.1, or C.P.2. In both cases the cosmological coincidence problem is solved~(see the $\Omega_\mathrm{de}/\Omega_\mathrm{dm}$ column of Table~\ref{tab:Table1}). On the other hand, if we consider $Q = 3 H b \rho_\mathrm{dm}$, then the following prior constraints $\omega_{0} \in [-1,1]$, $\omega_{1} \in [-1,1]$ and $b \geq 0$ follow, so that we obtain three stable critical solutions. The features of these solutions are presented in Table~\ref{tab:Table1_1}, the most interesting solution being the critical point C.P.4.

\begin{table}[t]
  \centering
    \begin{tabular}{ | l | l | l | l  | l | l | l | p{2cm} |}
    \hline
C.P. & $x$ & $y$ & $q$ & $\omega_\mathrm{de}$ & $\omega_\mathrm{eff}$ & $\Omega_\mathrm{de}/\Omega_\mathrm{dm}$ & Type of stability \\
      \hline
C.P.3 & $1$ & $-1$ & $-1$ & $-1$ & $-1$ & $--$ & Stable Node\\
          \hline
C.P.4 & $\frac{-2 b}{-3 b \omega_{1}+2 \omega_{0}+\omega_{1}}$ & $-b$ & $\frac{1-3 b}{2} $ & $\frac{-3 b \omega_{1}+2 \omega_{0}+\omega_{1}}{2}$ & $-b$ & $\frac{-2 b}{b (2-3 \omega_{1})+2 \omega_{0}+\omega_{1}}$ & Stable Node\\
     \hline
C.P.5 & $1$ & $\frac{2 \omega_{0}+\omega_{1}}{2-3 \omega_{1}}$ & $\frac{1}{2} \left(\frac{3 (2 \omega_{0}+\omega_{1})}{2-3 \omega_{1}}+1\right)$ & $\frac{2 \omega_{0}+\omega_{1}}{2-3 \omega_{1}}$ & $\frac{2 \omega_{0}+\omega_{1}}{2-3 \omega_{1}}$ & $--$ & Stable Node\\
   \hline

    \end{tabular}
    \vspace{5mm}
\caption{Two stable critical points corresponding to the interacting model with $Q = 3Hb\rho_\mathrm{dm}$. The type of stability has been analysed by imposing the following prior constraints: $\omega_{0} \in [-1,1]$, $\omega_{1} \in [-1,1]$, and $b \geq 0$.}
  \label{tab:Table1_1}
\end{table}

\begin{figure}[h!]
\begin{center}$
\begin{array}{cccc}
\includegraphics[width=80 mm]{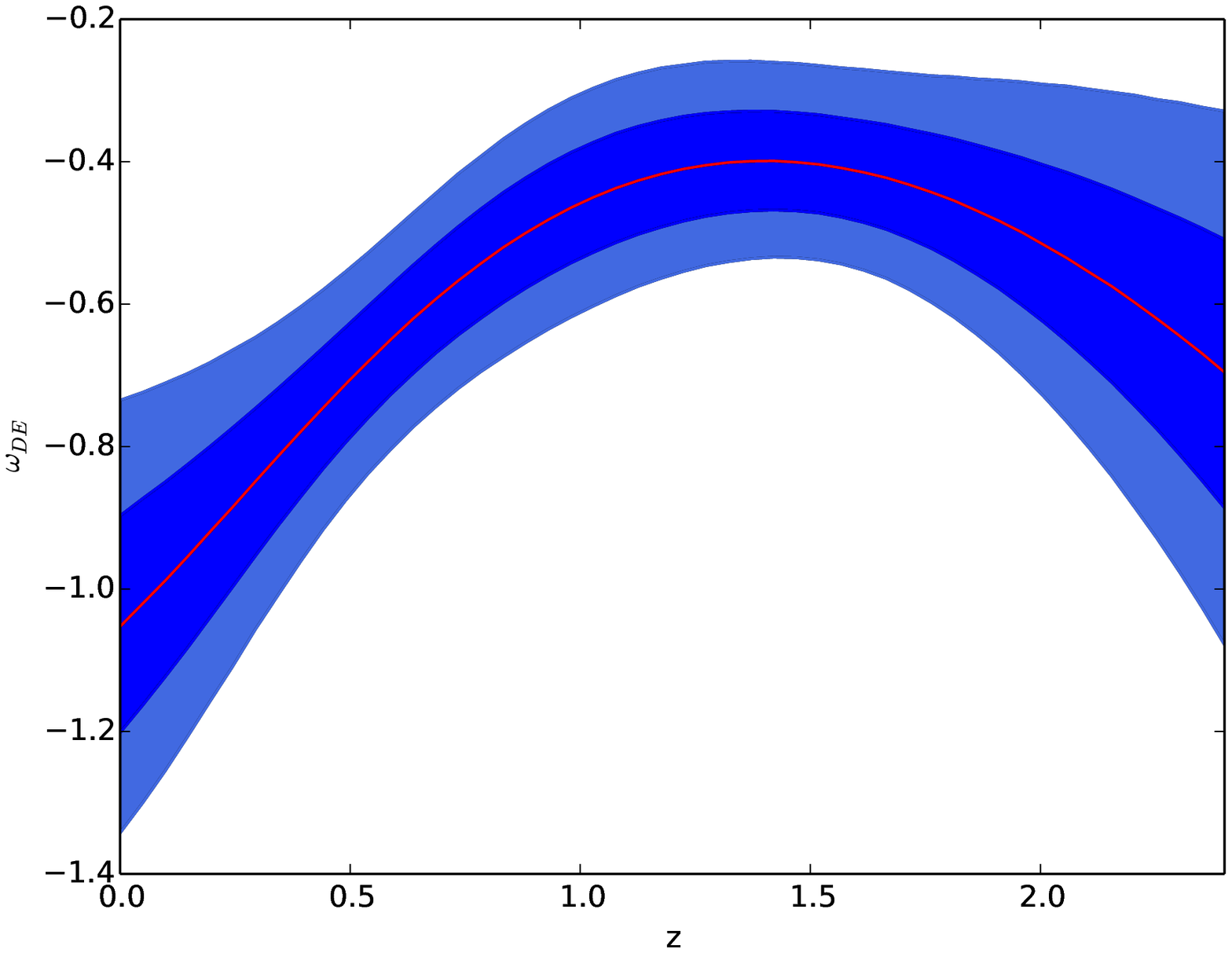}&
\includegraphics[width=80 mm]{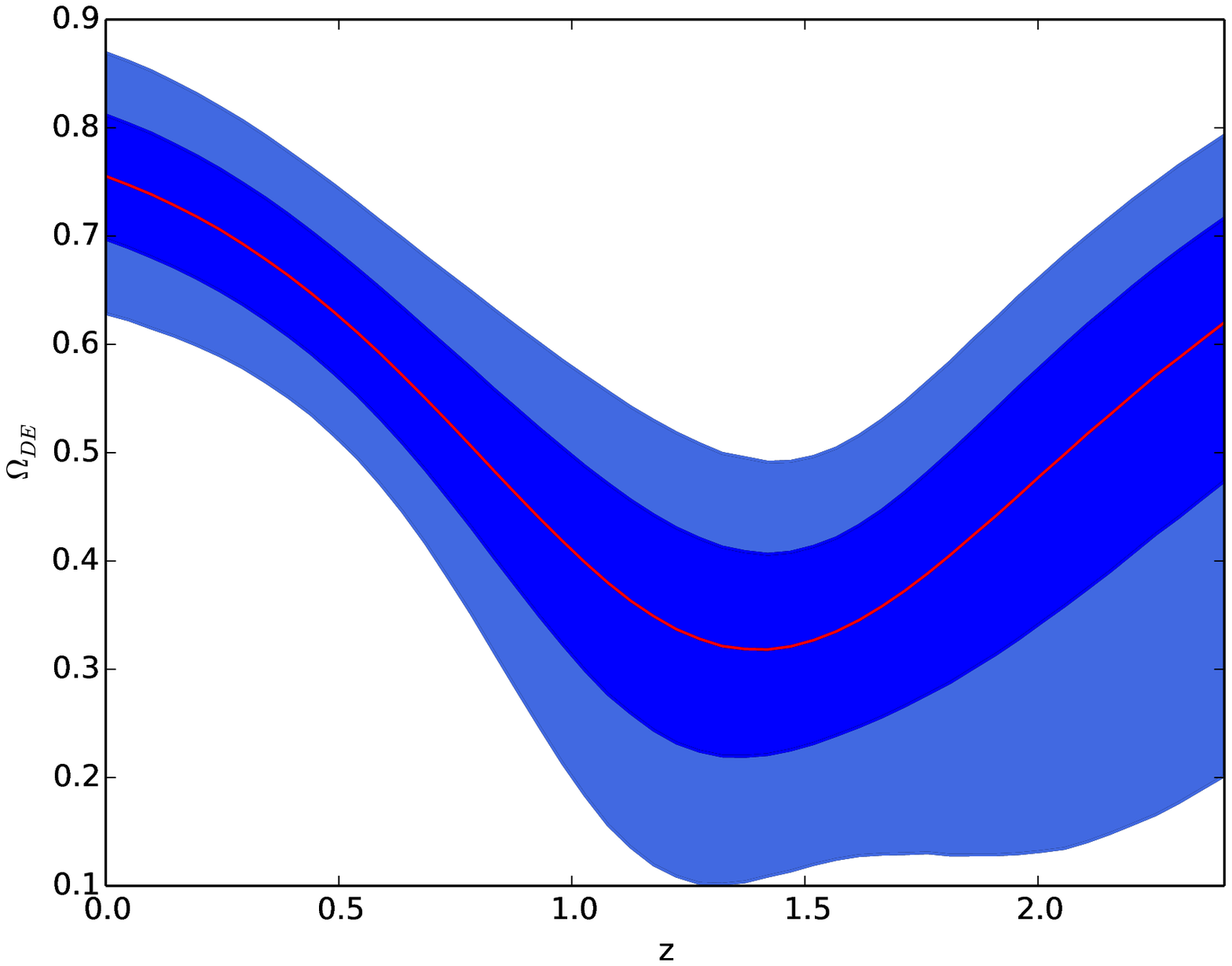}\\
\includegraphics[width=80 mm]{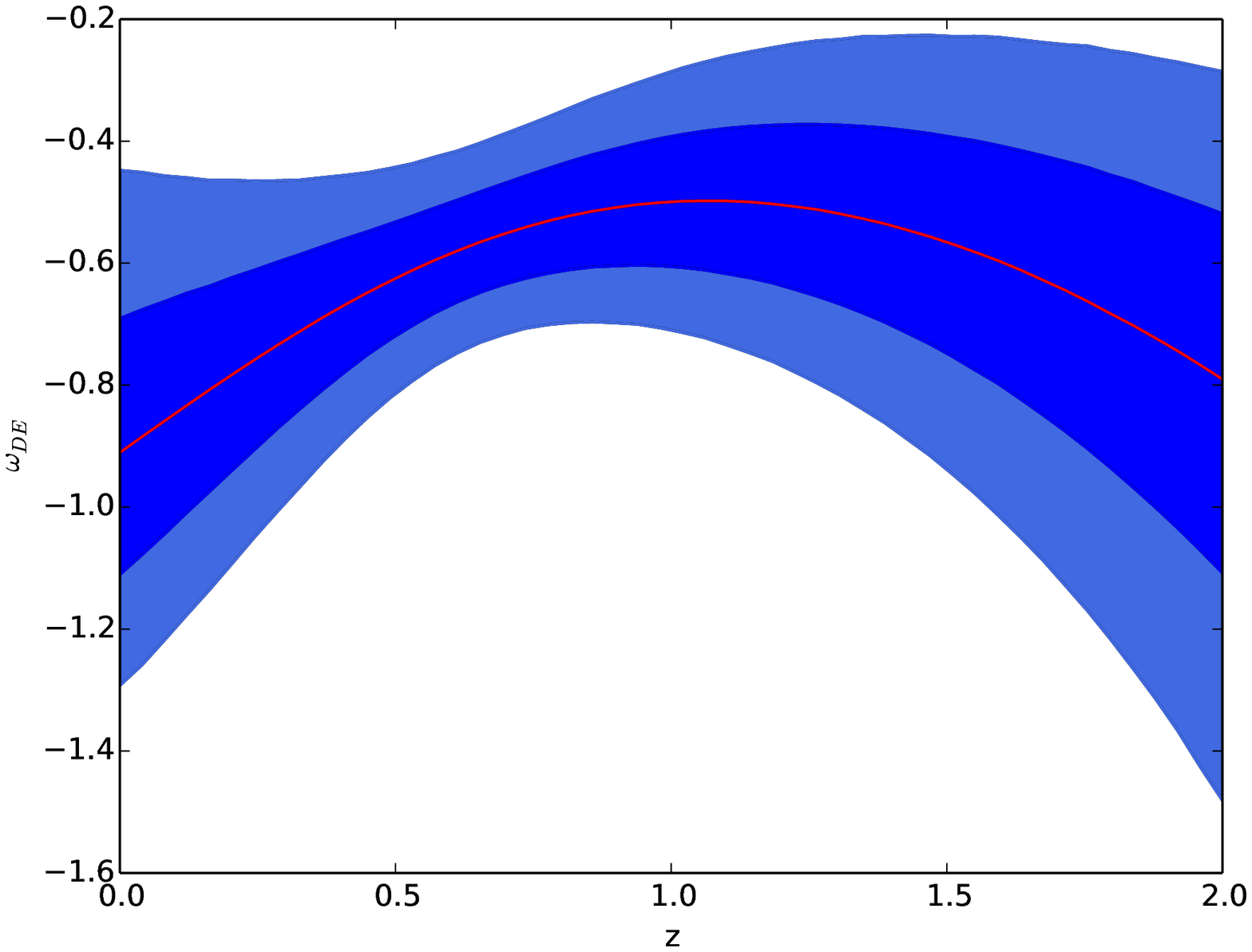}&
\includegraphics[width=80 mm]{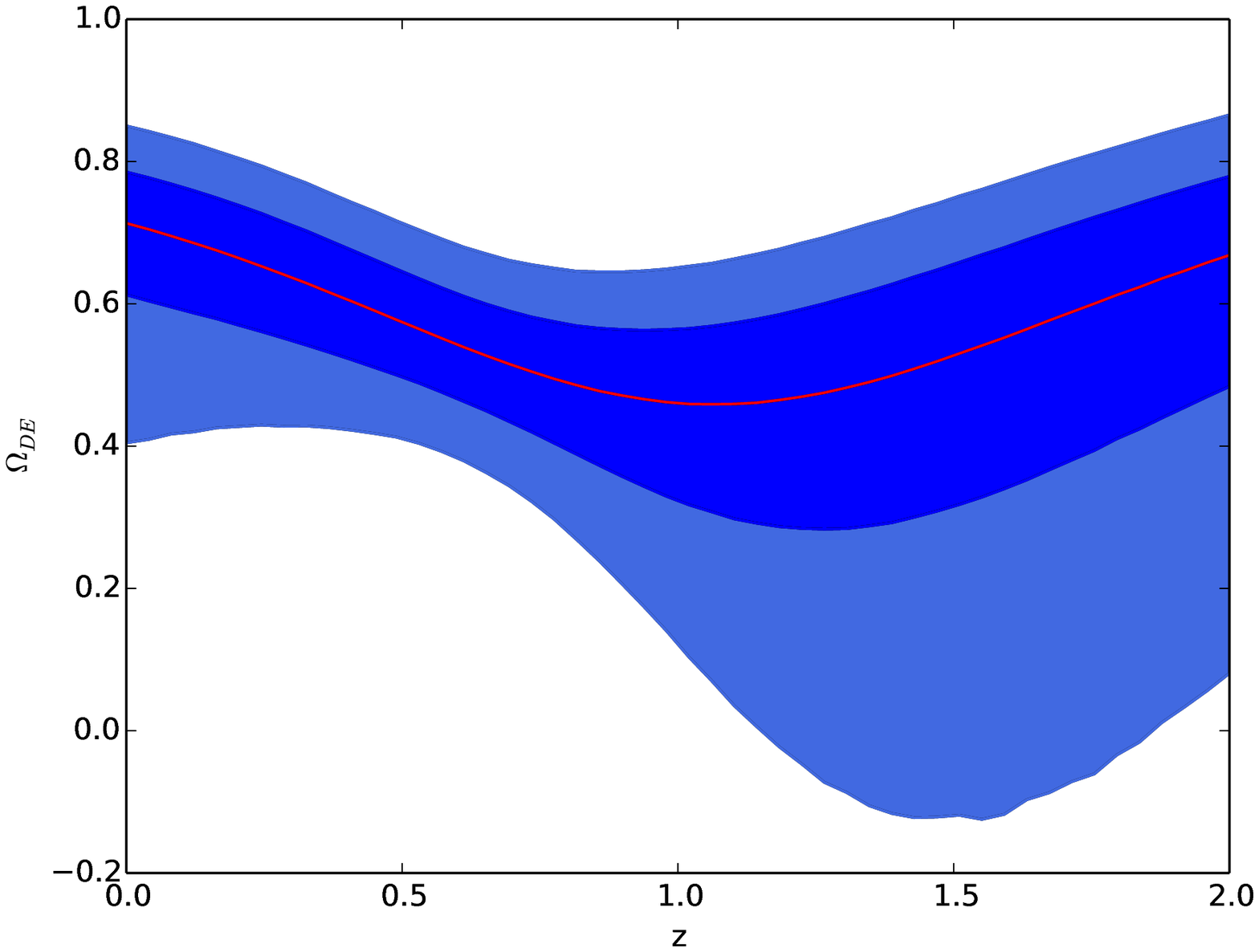}\\
\end{array}$
\end{center}
\caption{Reconstruction of the EoS, Eq.~(\ref{eq:omegaDE}), and $\Omega_\mathrm{de}$  parameters from the $H(z)$ data depicted in Table~\ref{tab:Table0}. The top panel represents GP reconstruction from $40$-point $H(z)$ data, while the bottom panel presents the GP reconstruction from $30$-point $H(z)$ data. The solid line is the mean of the reconstruction and the shaded blue regions are the $68\%$ and $95\%$ C.L. of the reconstruction, respectively.}
 \label{fig:d}
\end{figure}

Now, let us turn our attention on the study of the results obtained from the GP. In this case, first of all the main problem is to find appropriate constraints on the parameters $\omega_\mathrm{0}$ and $\omega_\mathrm{1}$  of Eq.~(\ref{eq:omegaDE}) EoS. While the second cosmological parameter which mainly can be interpreted as a controlling parameter needed in the analysis could be taken as $\Omega_\mathrm{de}$ at $z=0$~(this is due to structure of the equations). The reconstructed behavior of $\omega_\mathrm{de}$ and $\Omega_\mathrm{de} = \rho_\mathrm{de}/3H^{2}$ for both data combinations are presented in Fig.~\ref{fig:d}.  From the study of the bottom panel of  Fig.~\ref{fig:d} corresponding to the reconstruction obtained from the $30$-point sample data, we see that the $2\sigma$ errors due to GPs reconstruction do not exclude the possibility that $\Omega_\mathrm{de}<0$  for $z \in (1.1,1.9)$. However, from the $40$-point sample data we do not get the same conclusion~(see the top panel of Fig.~\ref{fig:d}). Moreover, in both cases $\omega_\mathrm{de}< 0$ i.e.  Eq.~(\ref{eq:omegaDE}) will describe dark energy without any transition to it from another kind of energy source. On the other hand, we should remember that the considered parametrization of the dark side of the Universe is valid for lower redshifts, therefore it is reasonable to expect that the reconstructed low redshift behavior of the parameters is correct and correspond to the recent Universe. We should highlight that this is correct for the parameters reconstructed using the reconstructed Hubble data, and not the Hubble data itself.  Now, having in mind the last point, let us study the behavior of the mean of the $\omega_\mathrm{de}$ parameter. In both cases we observe that the reconstructed mean indicates that the considered EoS at higher redshifts describes quintessence dark energy, however, an interesting situation could be observed for lower redshifts. In particular, with $40$-point data we observe the phantom divide crossing and that $\omega_\mathrm{de}$ at $z=0$ yields a value in well agreement with PLANCK 2015. On the other hand, $30$-point data reconstruction tells us that only quintessence dark energy will contribute into the background dynamics at lower redshifts. This is an excellent example demonstrating how the situation in general can be changed with the algorithm of data  filtering~(this filtering should be understood in a general term). Using the reconstructed Hubble parameter and its derivatives~(in this case only the first-order derivative), we reconstructed the deceleration parameter and thus the form of the EoS for the corresponding dark energy. Eventually, after a detailed study we come to the following constraints on  $\omega_\mathrm{0} = -0.6\pm 0.05$ and $\omega_\mathrm{1} = 0.65 \pm 0.05$, using the $40$-point sample of $H(z)$ data as described in Table~\ref{tab:Table0}. Moreover, in both cases at higher redshifts we should have a dark energy dominated Universe in order to explain the reconstructed accelerated expansion at higher redshifts~(see Fig.~(\ref{fig:c}) for details). We would like to mention also that the $40$-point sample of $H(z)$ data used in the presented analysis predicts a higher $\Omega_\mathrm{de}$ at $z=0$, as compared with the reconstruction from $30$-point samples of $H(z)$ data.

\begin{figure}[h!]
\begin{center}$
\begin{array}{cccc}
\includegraphics[width=80 mm]{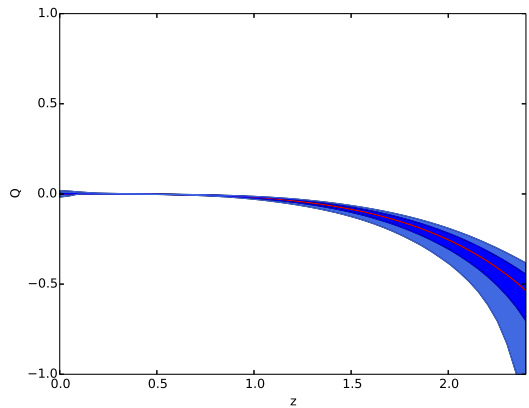}&
\end{array}$
\end{center}
\caption{Reconstruction of the non-gravitational interaction $Q$ between dark energy and dark matter. The plot represents the GP reconstruction obtained from $40$-point samples of $H(z)$ data. The solid line is the mean of the reconstruction and the shaded blue regions are the $68\%$ and $95\%$ C.L. bands of the reconstruction, respectively.}
 \label{fig:e}
\end{figure}

On the other hand, Fig.~(\ref{fig:e}) presents the expected behavior of the reconstructed interaction between dark energy and dark matter. As we can see, according to the considered parametrization, Eq.~(\ref{eq:omegaDE}), we have to introduce some flow from dark matter to dark energy for higher redshifts in order to explain the observed value of the Hubble parameter at $z=2.34$. Moreover, the introduced interaction will quickly disappear and only for very low redshifts can we expect to see its influence back into the background dynamics.

\subsection{Model with $\omega_\mathrm{de} = \omega_{0} + \omega_{1}q(1+z)^{\alpha}$}\label{subsec:CC_2}

A different form of $\omega_\mathrm{de}$, namely
\begin{equation}\label{eq:omegaDEGen}
\omega_\mathrm{de} = \omega_{0} + \omega_{1}q(1+z)^{\alpha},
\end{equation}
where $\alpha$, $\omega_{0}$, and $\omega_{1}$ are parameters of the model, leads to the second cosmological model of this paper. We see that when $\alpha = 0$ this  model  reduces to the first one. Moreover, the corresponding autonomous system
\begin{align}
y^{\prime} =&  \frac{dy}{dN} = \frac{(3 y+1) \omega_{1} (x \omega_{0} ((\alpha +3) x+3 y)+y (3 (x-1) y-\alpha  x))+2 \omega_{0} ((\alpha +3) x+3 y) (x \omega_{0}-y)}{x \omega_{1} (3 x \omega_{0}+1)}, \nonumber \\
x^{\prime} =& \frac{dx}{dN} = 3 (-1 + x) y,
\end{align}
has the following two critical points:
\begin{equation}\label{eq:X1Y1M2}
(x_{1},y_{1}) = \left( 1, \frac{A + 3 (\alpha +4) \omega_{0} \omega_{1}+2 (\alpha +3) \omega_{0}+\alpha \omega_{1}-6 \omega_{0}^2}{6 (\alpha \omega_{1}+\omega_{0} (2-3 \omega_{1}))}\right ),
\end{equation}
and
\begin{equation}\label{eq:X2Y2M2}
(x_{2},y_{2}) = \left( 1, \frac{- A + 3 (\alpha +4) \omega_{0} \omega_{1}+2 (\alpha +3) \omega_{0}+\alpha \omega_{1}-6 \omega_{0}^2}{6 (\alpha \omega_{1}+\omega_{0} (2-3 \omega_{1}))}\right ),
\end{equation}
where $A = \sqrt{4 \omega_{0} \omega_{1} \left(3 \left(\alpha ^2-6\right) \omega_{0}-9 (\alpha +2)\omega_{0}^2+\alpha  (\alpha +3)\right)+4 \omega_{0}^2 (\alpha +3 \omega_{0}+3)^2+\omega_{1}^2 (3 (\alpha +2) \omega_{0}+\alpha )^2}$.

\begin{figure}[h!]
 \begin{center}$
 \begin{array}{cccc}
\includegraphics[width=85 mm]{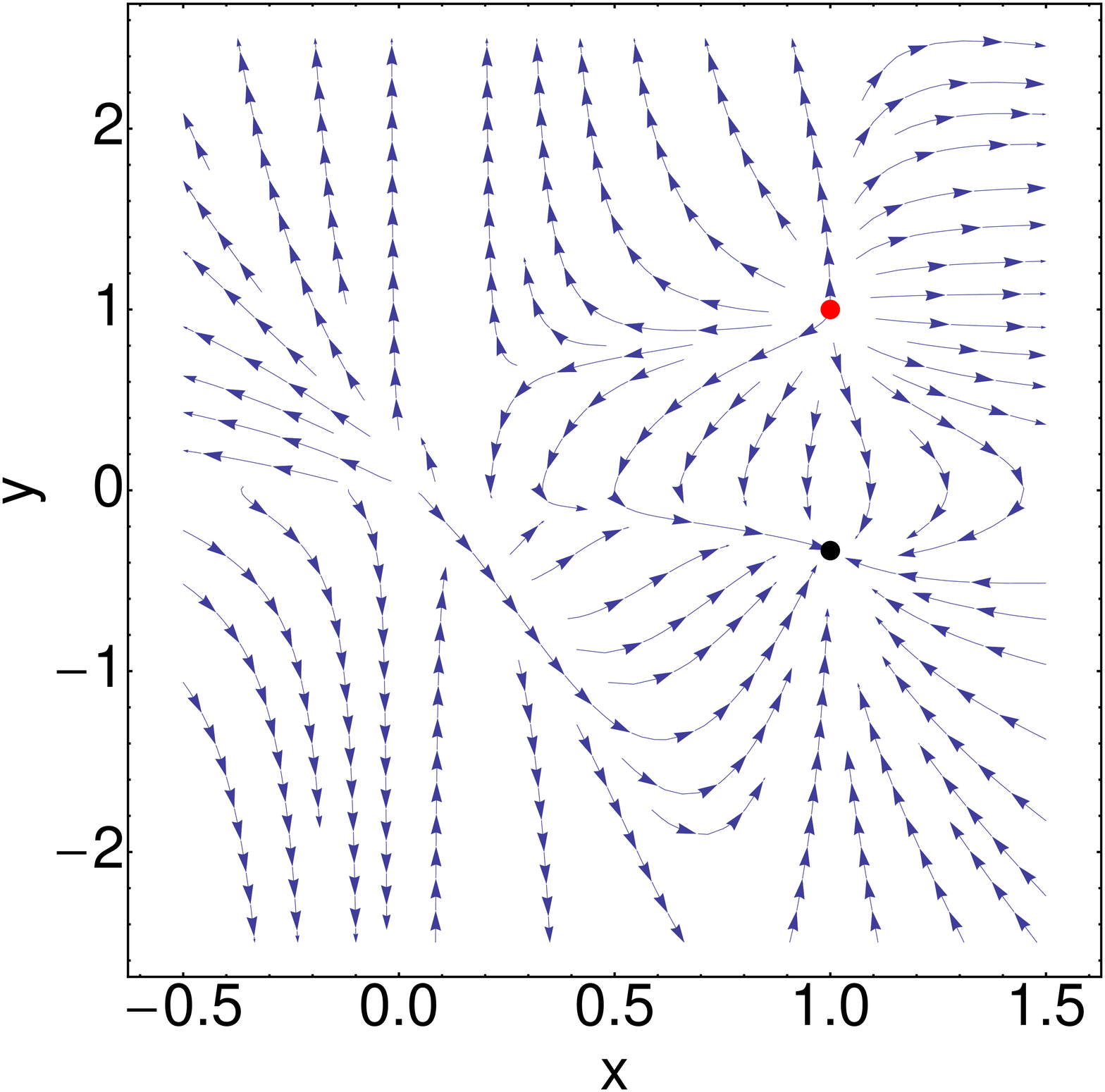} &
\includegraphics[width=85 mm]{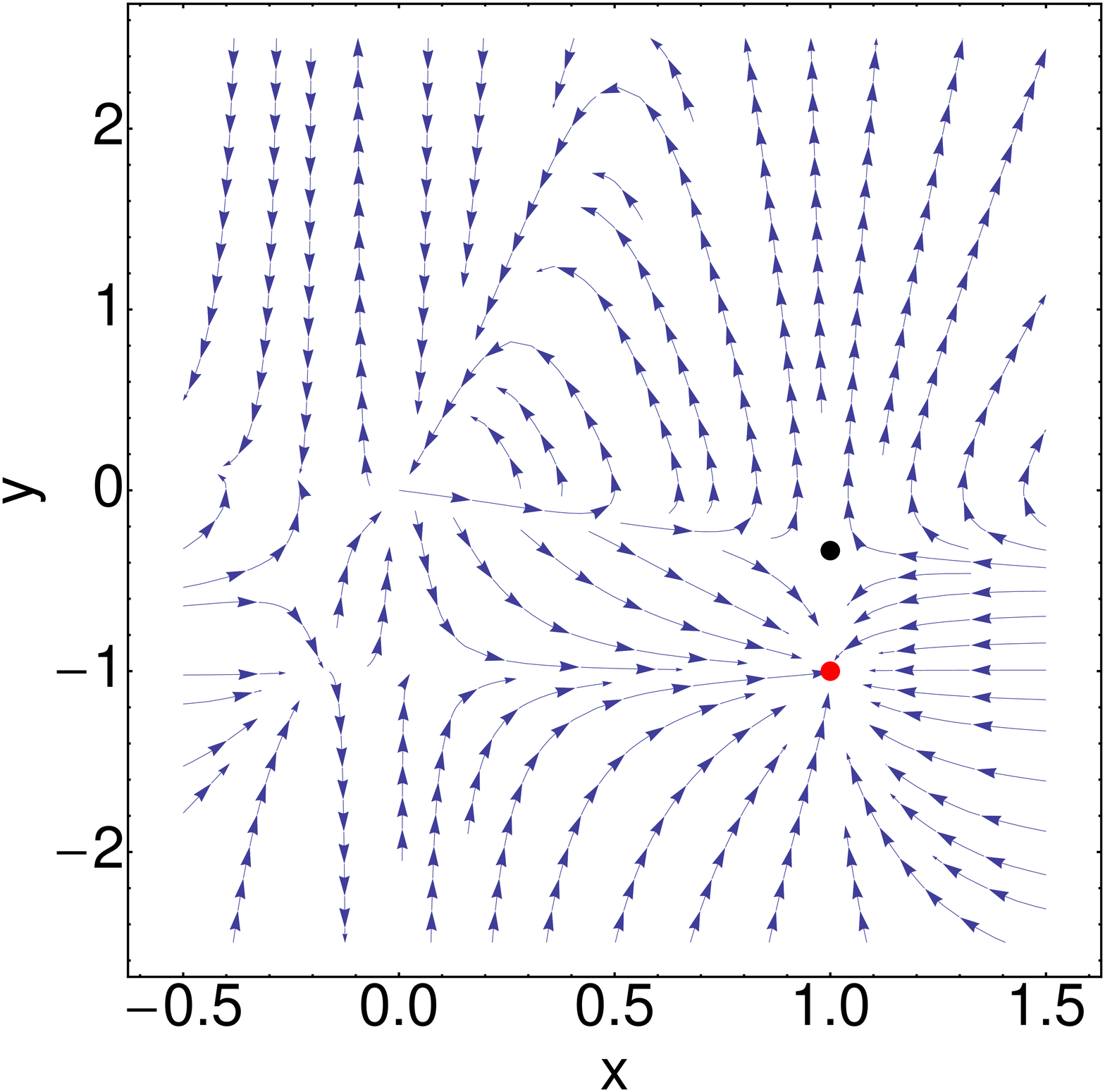}\\
\includegraphics[width=85 mm]{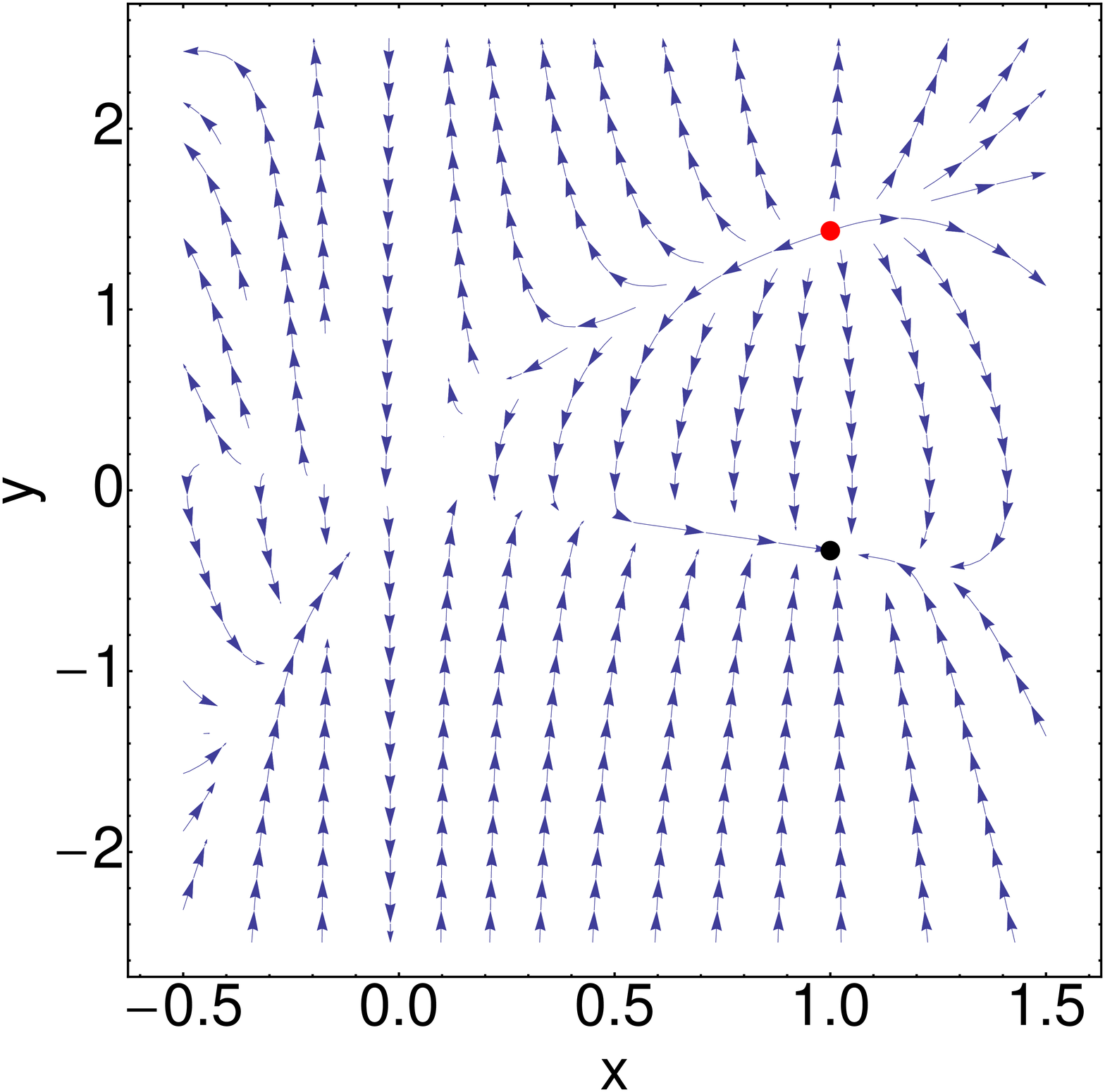} &
\includegraphics[width=85 mm]{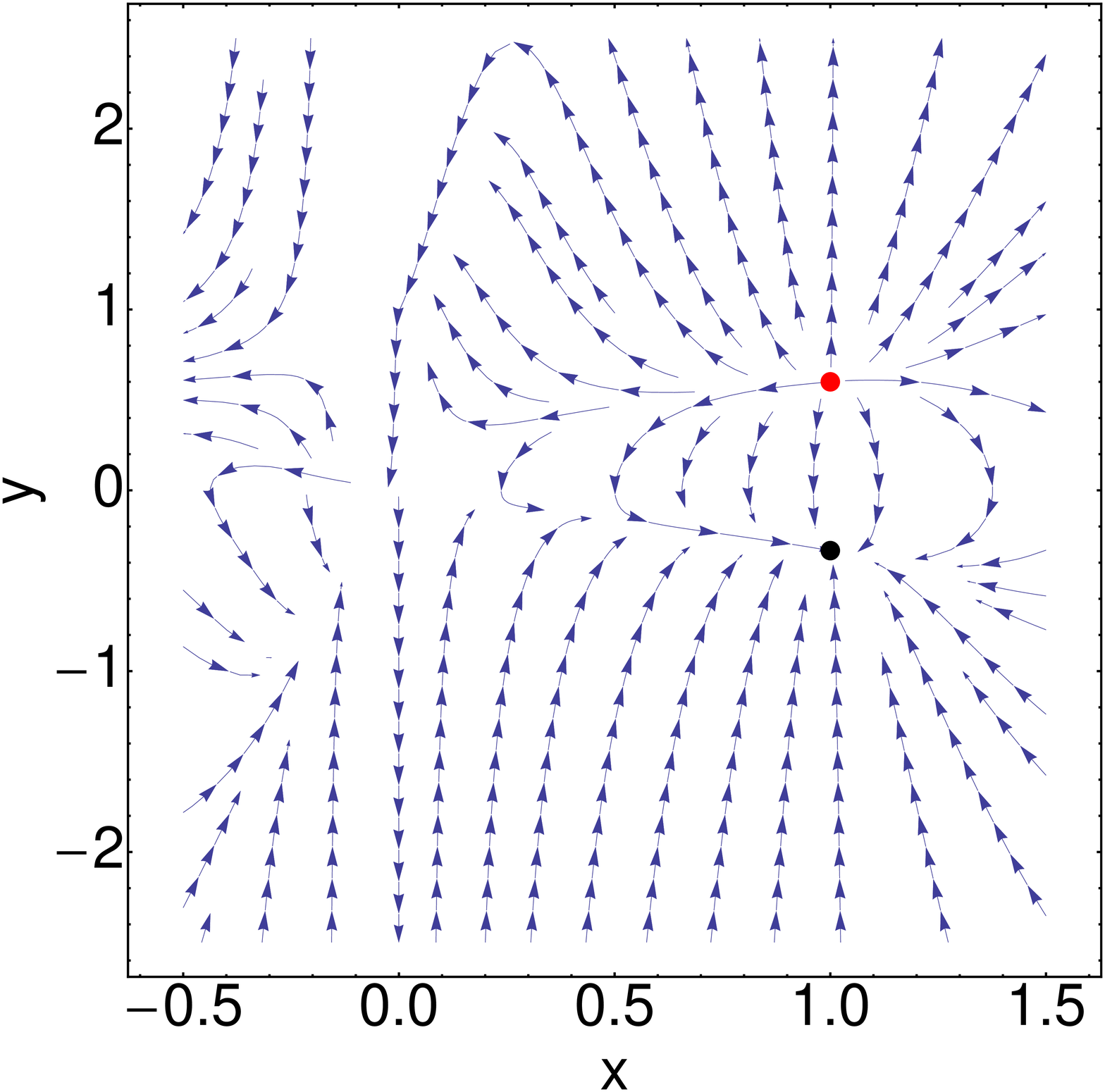}\\
 \end{array}$
 \end{center}
\caption{Phase space plot for the cosmological model with dark fluid described by Eq.~(\ref{eq:omegaDEGen}). The red dots represent the solutions corresponding to Eq.~(\ref{eq:X1Y1M2}), while those corresponding to Eq.~(\ref{eq:X2Y2M2}), for various values of $\omega_{0}$, $\omega_{1}$, and $\alpha$, are depicted in black dots. We have set  $\omega_{0} = -2.0$, $\omega_{1} = 2.0$, and $\alpha=-2$, in the case of the top-left plot; $\omega_{0} = 2.0$, $\omega_{1} = 2.0$ and $\alpha=-2$ for the top-right plot;
 $\omega_{0}=3.0$, $\omega_{1}=-0.5$, and $\alpha = -2$ for the left-bottom one; and $\omega_{0}=3.0$, $\omega_{1}=-1.5$ with $\alpha = -2$ for the right-bottom plot. Remember that $x\in [0,1]$.}
 \label{fig:0_2}
\end{figure}

Note that both solutions describe the state of the Universe with background dynamics governed only by the dark energy considered, in each case. What is more, both solutions are stable nodes with respect to the constrains considered  previously for the non-interacting model. In this case, the parameter space can be divided into appropriate regions and  the solutions can be stabilized separately. The phase plots  in Fig.~\ref{fig:0_2} clearly show, in four different cases, how one of the solutions is an attractor, while the other remains an unstable critical point. In particular, from the top panel of Fig.~\ref{fig:0_2} we see, that in the case of the left plot corresponding to $\omega_{0}=-2.0$, $\omega_{1}=2.0$ and $\alpha = -2$,  the critical point Eq.~(\ref{eq:X1Y1M2}) is an unstable node, while the critical point Eq.~(\ref{eq:X2Y2M2}) is a stable one. On the other hand, the phase plot depicted on the right hand side shows that the critical point Eq.~(\ref{eq:X1Y1M2}) is a stable node, while the critical point Eq.~(\ref{eq:X2Y2M2}) is a saddle point~(in this case we have set $\omega_{0} = 2.0$, $\omega_{1} = 2.0$ and $\alpha=-2$). The two other plots at the bottom exhibit the unstable node nature of the critical point of Eq.~(\ref{eq:X1Y1M2})~(for $\omega_{0}=3.0$, $\omega_{1}=-0.5$, and $\alpha = -2$) and the stable node nature of the critical point of Eq.~(\ref{eq:X2Y2M2})~(for $\omega_{0}=3.0$, $\omega_{1}=-1.5$, and $\alpha = -2$). The study carried out in this section corresponds to the non-interacting dark energy models~(except for the phase space analysis of the first model), and proves the viability of the models under discussion to explain the accelerated expansion of the Universe at large scale.

\begin{figure}[h!]
\begin{center}$
\begin{array}{cccc}
\includegraphics[width=80 mm]{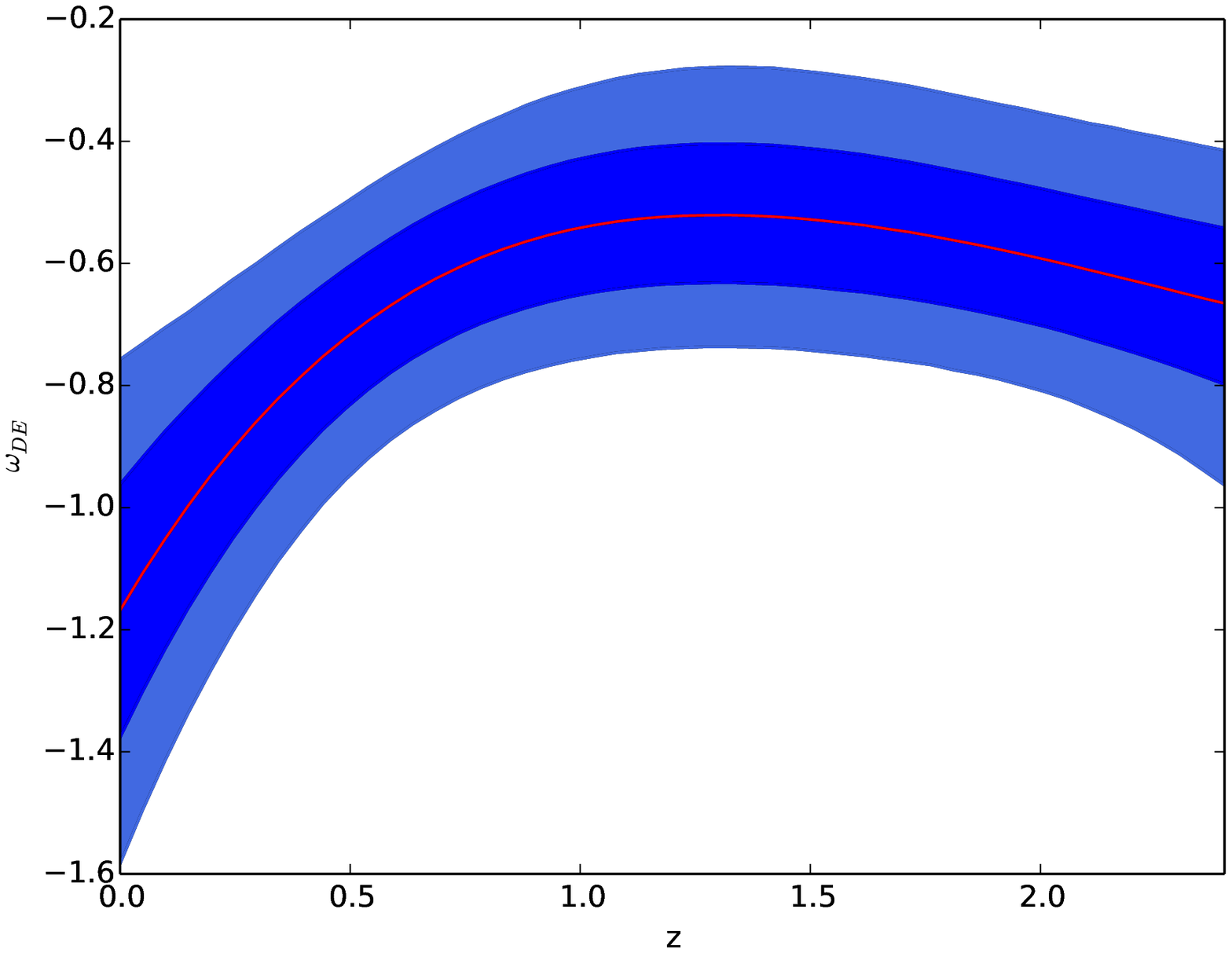}&
\includegraphics[width=80 mm]{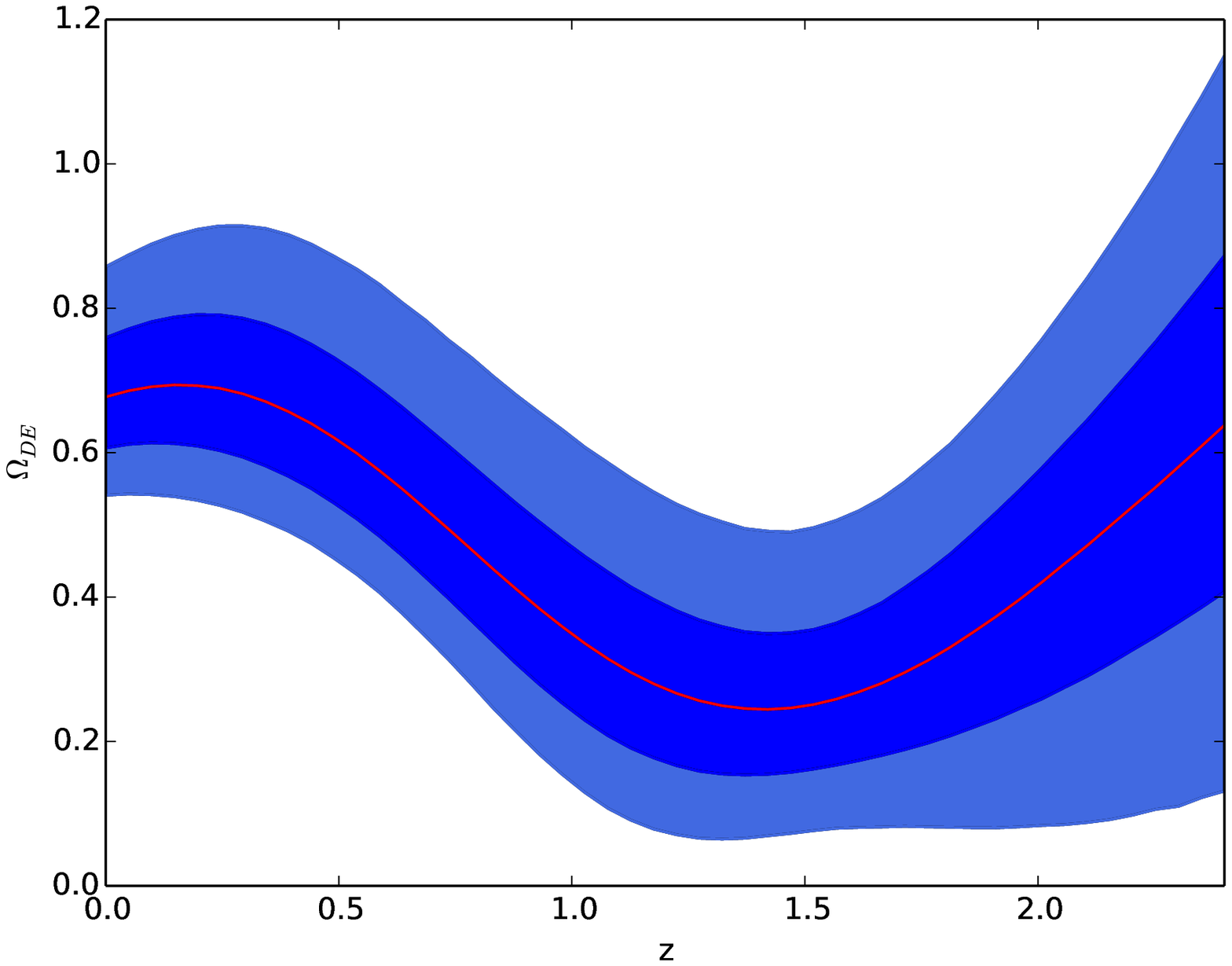}\\
\includegraphics[width=80 mm]{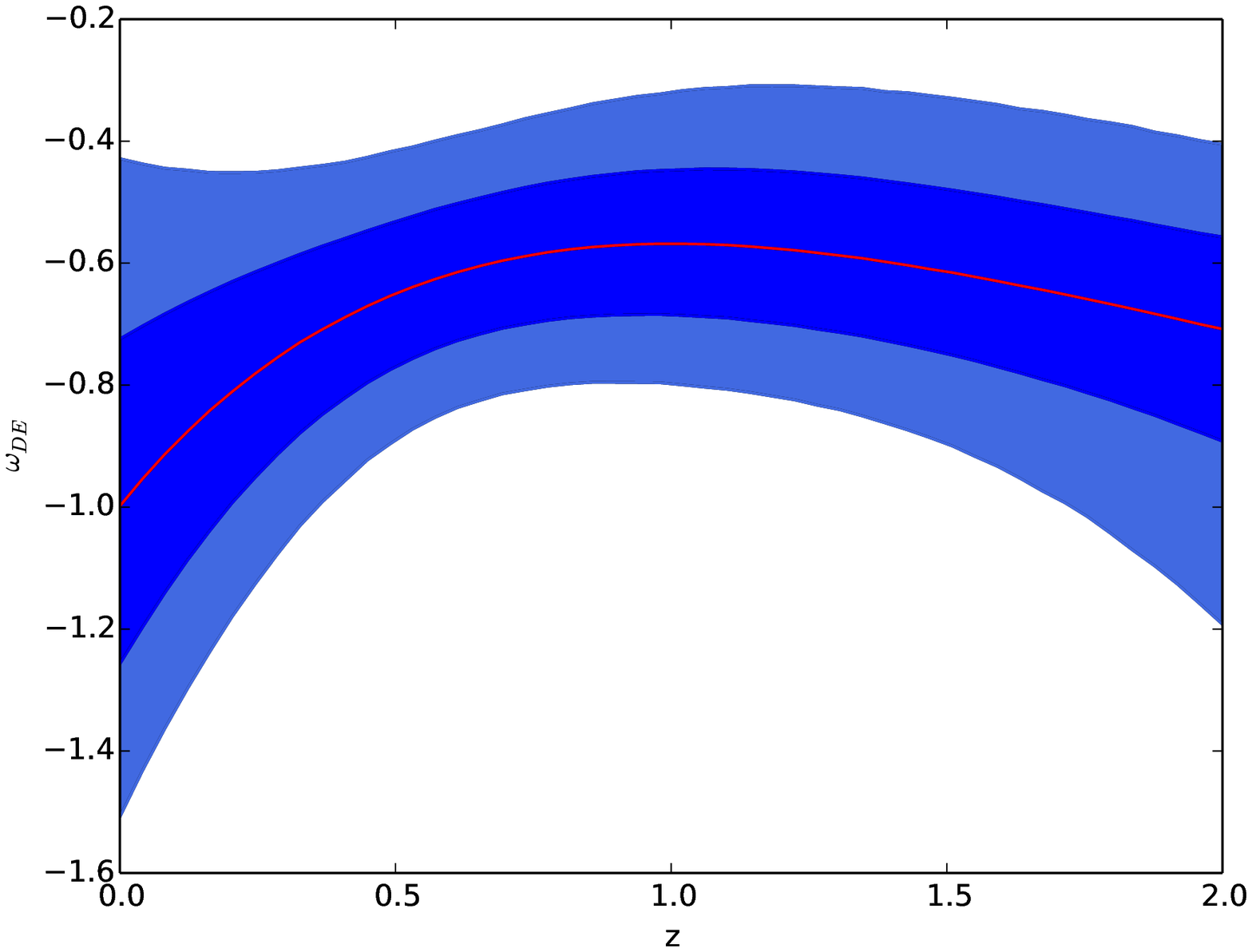}&
\includegraphics[width=80 mm]{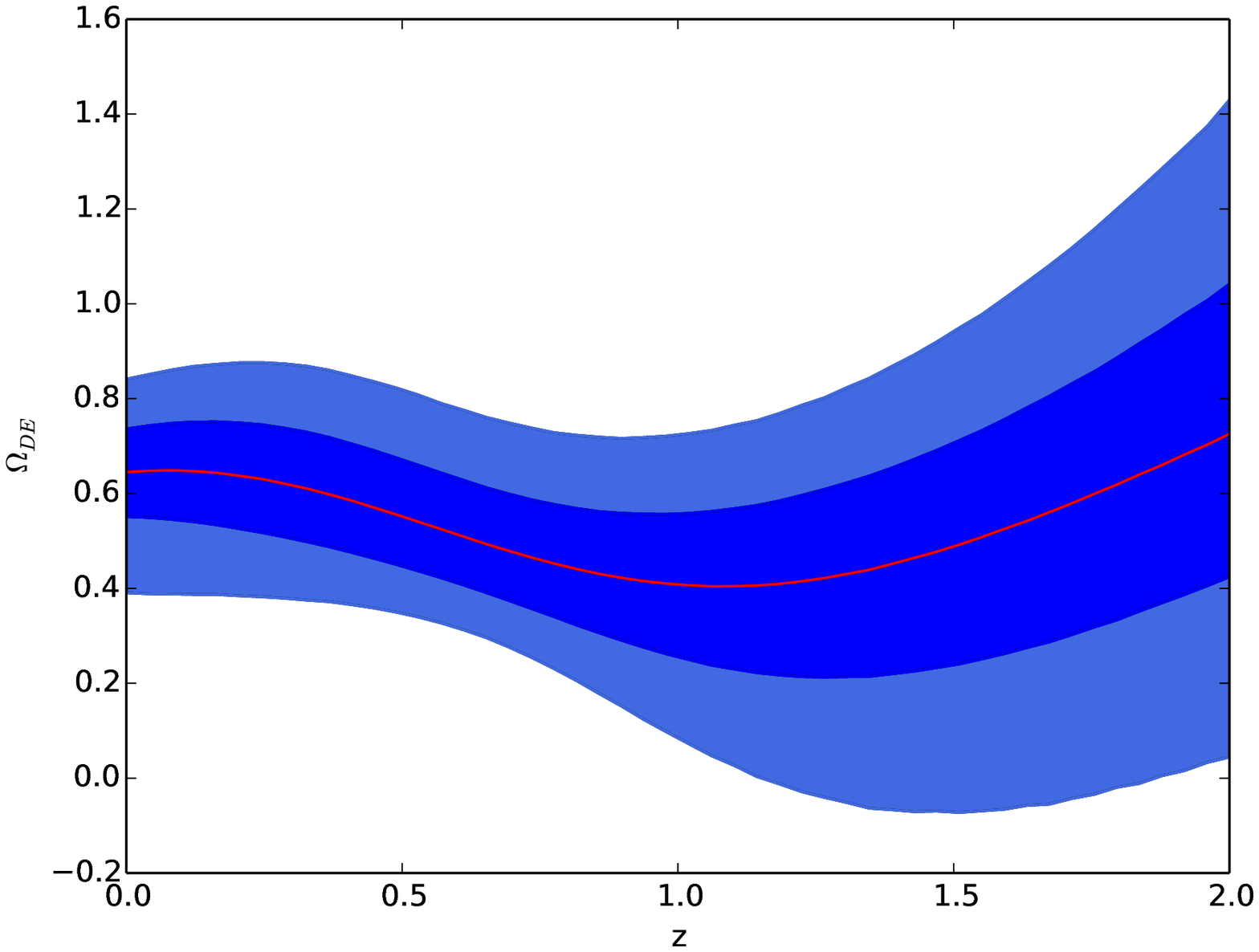}\\
\end{array}$
\end{center}
\caption{Reconstruction of the EoS, Eq.~(\ref{eq:omegaDEGen}), and $\Omega_\mathrm{de}$  parameters from the $H(z)$ data presented in Table~\ref{tab:Table0}. The top panel represents the GP reconstruction from $40$-point samples of $H(z)$ data, while the bottom panel presents the GP reconstruction from $30$point samples of $H(z)$ data. The solid line is the mean of the reconstruction and the shaded blue regions are the $68\%$ and $95\%$ C.L. bands of the reconstruction, respectively. In this case we fixed $\omega_\mathrm{0} = -0.63 \pm 0.1$, $\omega_\mathrm{1} = 0.78 \pm 1$ and $\alpha = -1 \pm 0.5$, while the value of the Hubble parameter $H_\mathrm{0}$ at $z=0$ has not been fixed.}
 \label{fig:f}
\end{figure}

Now, let us discuss the results corresponding to the GP reconstruction for this model. First of all, similar to the model with $\alpha = 0$, we see that according to the $95\%$ C.L. of the reconstruction, the possibility that $\Omega_\mathrm{de}<0$ at higher redshifts is not excluded~(if we take into account the $30$-point sample of $H(z)$ data). On the contrary, consideration of the $40$-point sample of $H(z)$ data reconstruction shows that the model can be viable. It should be mentioned also that, in this case, similar to the model with $\alpha = 0$, we observed that to have $\Omega_\mathrm{de} < 0.7$ at $z=0$ always needs a phantom dark energy~(this can be seen on the top panel of Fig.~\ref{fig:f}). However, the reconstruction based on $30$-point samples of $H(z)$ data shows that for the same values of the parameters, as in the case discussed above, instead of having phantom dark energy at $z=0$, we will have the cosmological constant with $\omega_\mathrm{de} = -1$ giving $\Omega_\mathrm{de} = 0.672$. In general the study shows that the $H(z)$ data used supports mainly negative values of the parameter $\alpha$. Moreover, the study of the reconstructed form of the non-gravitational interaction in this case also shows that at higher redshifts we had to introduce a rapidly vanishing non-gravitational interaction providing energy transfer from dark matter to dark energy in order to explain the value of the Hubble parameter at $z=2.34$. On the other hand, the parameter $\alpha$ does not affect the form and the behavior of the reconstruction of the interaction term. This could mean that the $\alpha$ parameter will not affect  the type of future singularities indicated by the existence of non-gravitational interaction. A similar picture has been observed also with the analysis of numerical solutions directly obtained from the field equations for appropriate initial conditions. Now, we will try to use the results here obtained  and see if one should accept or reject the considered models. It turns out that the $H(z)$ data used is not enough to reach a final conclusion; however,  we should highlight that
\begin{enumerate}
\item The model, Eq.~(\ref{eq:omegaDE}), according $68\%$ C.L. of the reconstruction with $40$ sample $H(z)$ data can be accepted.
\item The model, Eq.~(\ref{eq:omegaDE}), according $68\%$ C.L. of the reconstruction with $30$ sample $H(z)$ data can be accepted.
\item The model, Eq.~(\ref{eq:omegaDEGen}), according $68\%$ C.L. of the reconstruction with $30$ sample $H(z)$ data can be accepted.
\end{enumerate}

\begin{figure}[h!]
\begin{center}$
\begin{array}{cccc}
\includegraphics[width=80 mm]{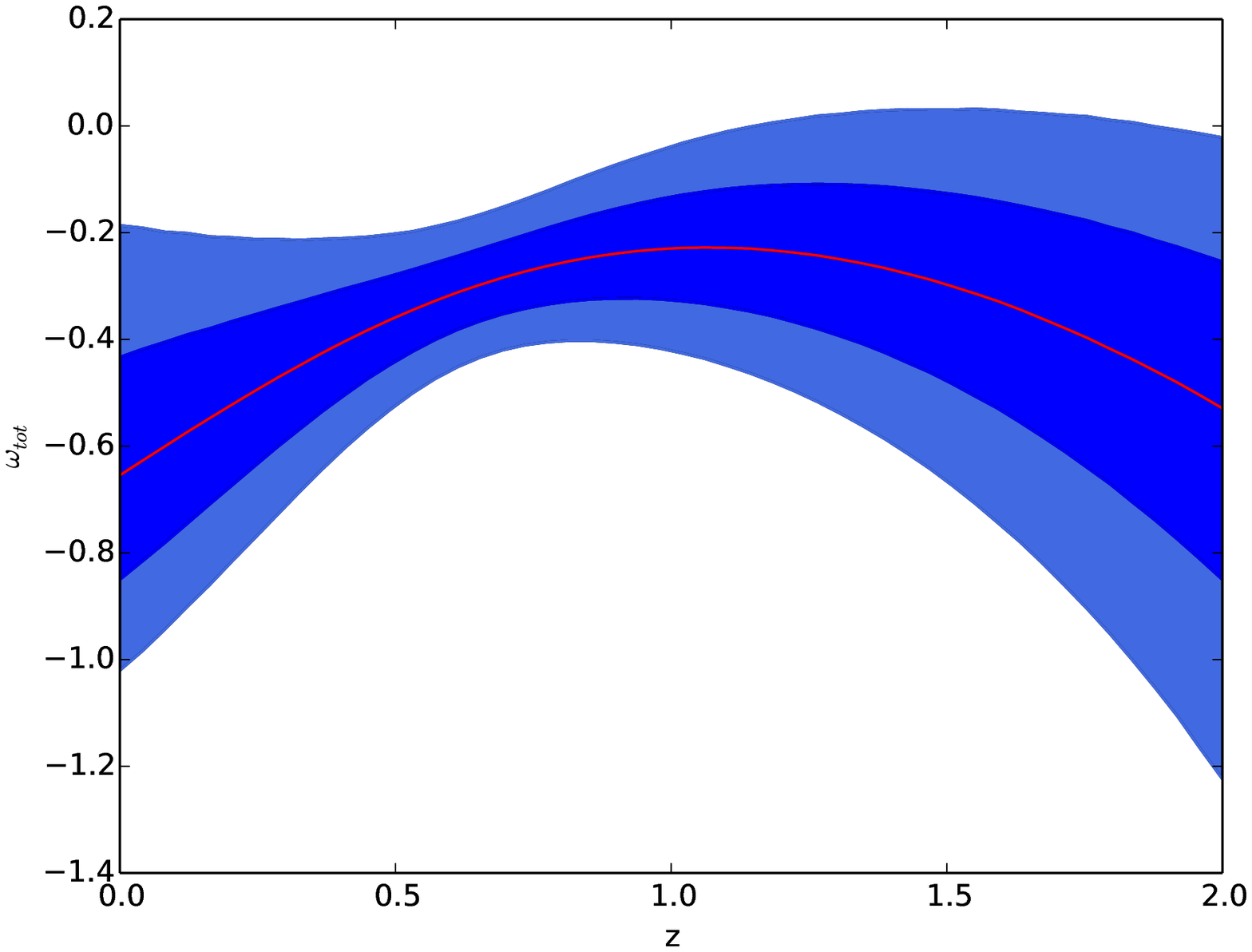}&
\includegraphics[width=80 mm]{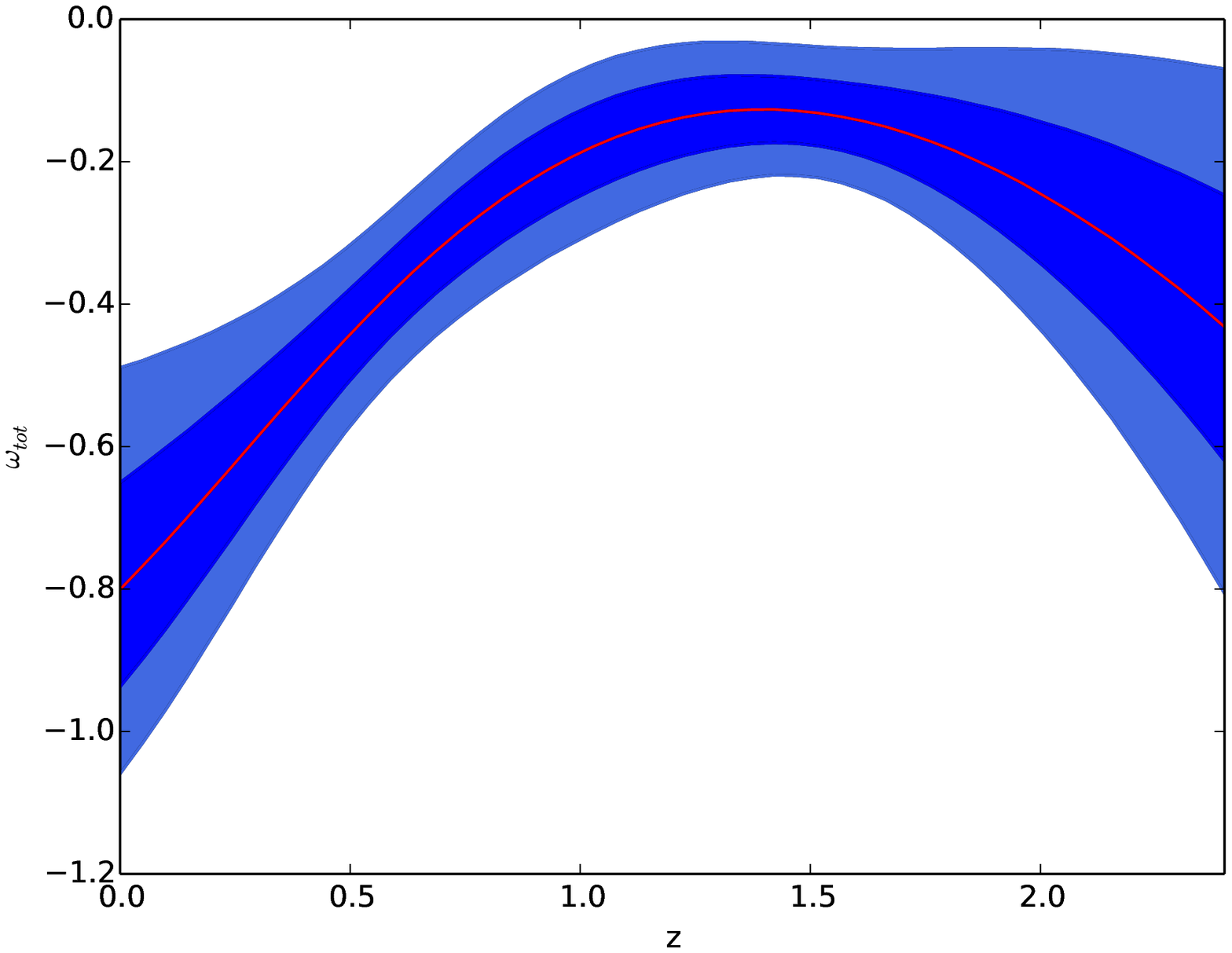}\\
\includegraphics[width=80 mm]{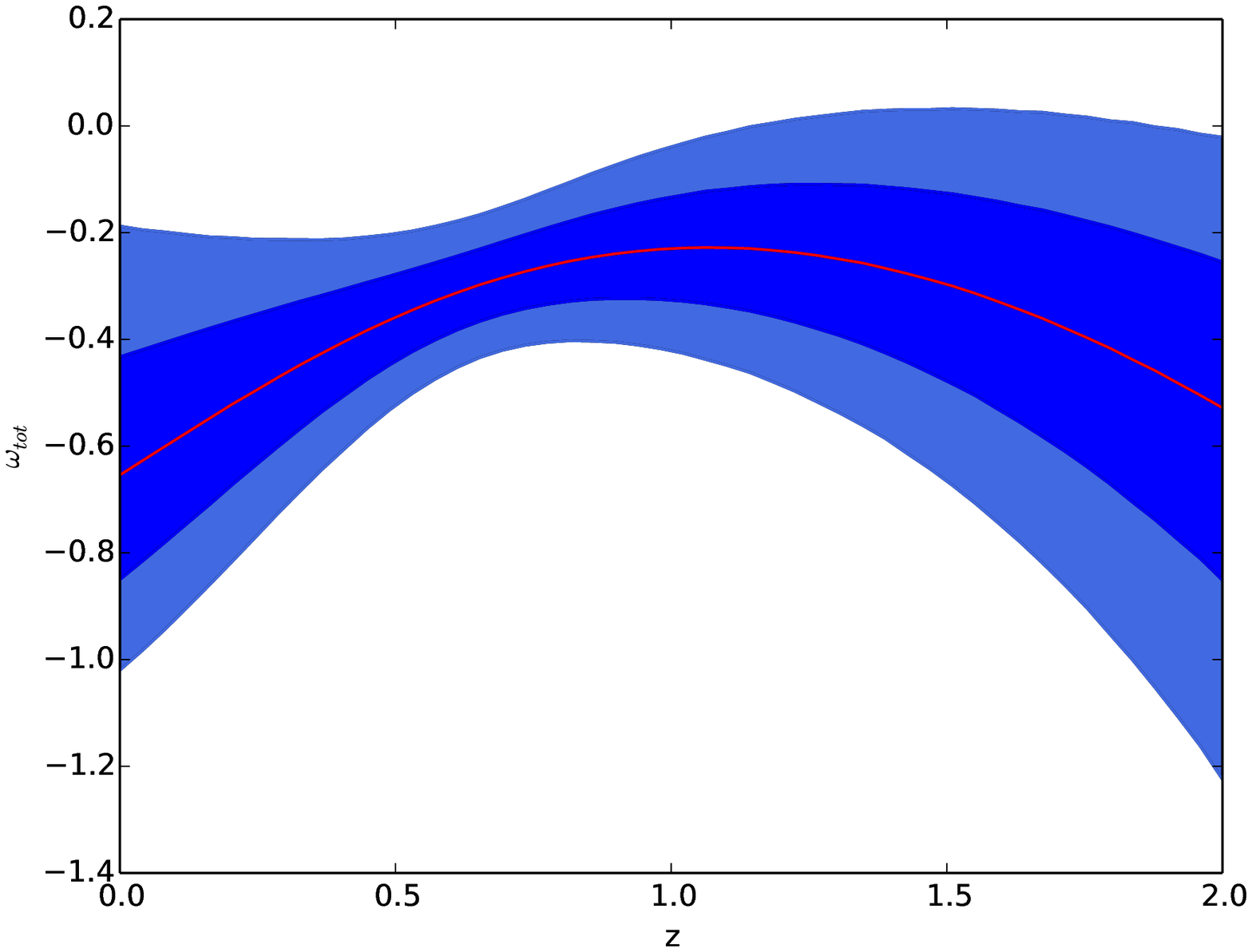}&
\includegraphics[width=80 mm]{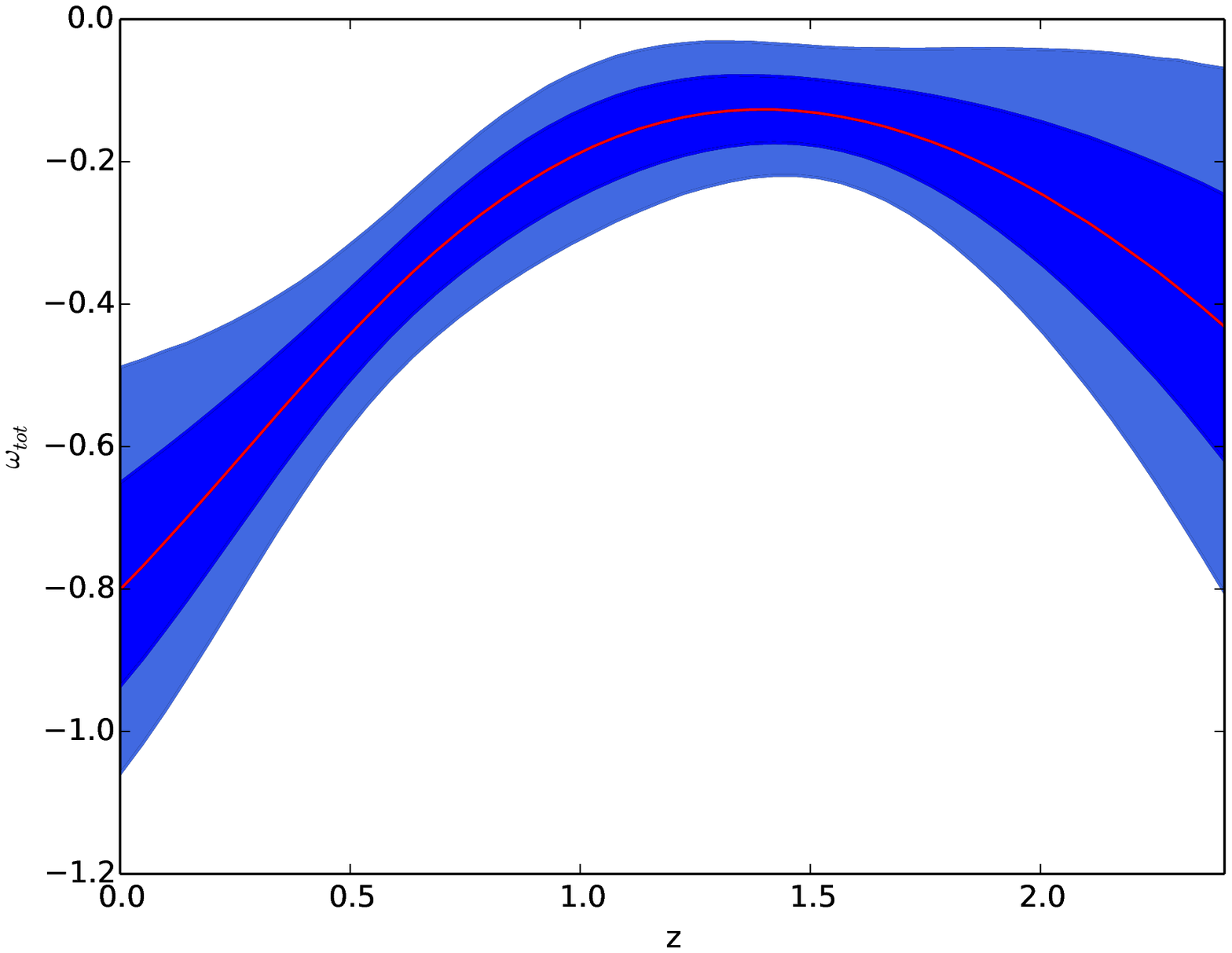}\\
\end{array}$
\end{center}
\caption{Reconstruction of the EoS parameter of the effective fluid from the $H(z)$ data depicted in Table~\ref{tab:Table0}. The top panel represents GP reconstruction from  $30$-  and $40$-point samples of $H(z)$ data, for the model given by Eq.~(\ref{eq:omegaDE}), respectively. On the other hand, the bottom panel presents the GP reconstruction from $30$- and $40$-point samples of $H(z)$ data for the model described by Eq.~(\ref{eq:omegaDEGen}). The solid line is the mean of the reconstruction and the shaded blue regions are the $68\%$ and $95\%$ C.L. of the reconstruction, respectively. In this case we fixed $\omega_\mathrm{0} = -0.63 \pm 0.1$, $\omega_\mathrm{1} = 0.78 \pm 1$, and $\alpha = -1 \pm 0.5$ and the value of the Hubble parameter $H_\mathrm{0}$ at $z=0$ has not been fixed.}
 \label{fig:g}
\end{figure}

As a last point of the analysis, let us investigate the behavior of the EoS parameter of the effective fluid for both models, as defined in the following way~( see Fig.~\ref{fig:g} )
\begin{equation}
\omega_\mathrm{eff} = \frac{P_\mathrm{de}}{\rho_\mathrm{de} + \rho_\mathrm{dm}}.
\end{equation}

It is easy to see that both models, according to the $68\%$ C.L. of the reconstruction, describe a quintessence dark fluid with $\omega_\mathrm{eff} > -1.0$. Moreover, if we will consider the $\omega_\mathrm{eff}$ parameter to decide acceptance or rejection of the model, then from Fig.~\ref{fig:g} we conclude that both models can be accepted even with the $95\%$ C.L. of the reconstruction. However, the above analysis shows another picture. The construction presented in this section for the second model, Eq.~(\ref{eq:omegaDEGen}), has been done with $\omega_\mathrm{0} = -0.63 \pm 0.1$, $\omega_\mathrm{1} = 0.78 \pm 1$ and $\alpha = -1 \pm 0.5$ and the value of the Hubble parameter $H_\mathrm{0}$ at $z=0$ has not been fixed.

Needless to say, the presented analysis should be extended involving different forms of covariance functions, as it is described in the literature. Moreover, it is very important to study the impact of different combinations of data sets  during reconstruction. In general GaPP is a very well organized publicly available Python code allowing to carry out a quick study of the mentioned questions~(at the algorithmic level, at least). We expect to report on our study of the mentioned points in a forthcoming paper.

\section{Some remarks on future singularities}\label{sec:TS}

It is well known that non-gravitational interactions of a certain type can have a significant effect on the background dynamics and structure formation processes; therefore, one can assume that this can affect the formation of the singularities in cosmological models. On the other hand, if a non-gravitational interaction can change the behavior of the EoS parameter, then with high probability within the considered model, it can change the type of singularity. The purpose of this section is to classify singularities assuming explicitly given forms of non-gravitational interactions. Unfortunately, GPs cannot be used to decide on aspects related to such future singularities. It is due to the fact that the covariance function is infinitely differentiable that quantities reconstructed  from the observational data cannot carry information about the future singularities either. If we want to assume that the reconstructed future behavior of the parameters based on the known data is totally reliable. This means that, eventually, we will have to get back and solve the field equations for appropriate initial conditions.

Now let us start the analysis of the numerical results corresponding to the first model with Eq.~(\ref{eq:omegaDE}) EoS. We observe, that the model can explain the accelerated expansion of the recent Universe and that the model is free from the cosmological coincidence problem. On the other hand, for instance, if we choose the following values of the parameters of the model to be as follows: $H_{0} = 0.676$, $ \Omega_\mathrm{dm}= 0.269$, $\omega_{0}=-0.985$ and $\omega_{1} = 0.02$, then we can obtain a good agreement with the constraints coming from the $Omh^{2}$ analysis~(based on the estimation of the $Omh^{2}$ parameter from $3$ $H(z)$ measurements obtained from BAO). An interesting recent discussion on this analysis can be found in Ref.~\cite{Zheng:2016jlq}, where a larger sample of $H(z)$ comprising $6$ BAO measurements and $23$ data points from cosmic chronometers~(differential ages of passively evolving galaxies) has been used. $Omh^{2}$ analysis is a modification of the $Om$ analysis discussed in Ref.~\cite{VarunSahni:2014} and it is defined in the following way
\begin{equation}\label{eq:OMh2}
Omh^{2}(z_{i};z_{j}) = \frac{h^{2}(z_{i}) - h^{2}(z_{j})}{(1+z_{i})^{3} - (1+z_{j})^{3}},
\end{equation}
where $h = H(z)/100$ km s$^{-1}$ Mpc$^{-1}$. The model independent $Om$ analysis, which conveniently allows to study different features of cosmological models itself is also very valuable analysis to explore the phase space  of a cosmological model. It is defined in the following way~\cite{VarunSahni:2008}
\begin{equation}\label{eq:OM}
Om = \frac{\tilde{h}^{2}(z) - 1}{(1+z)^{3} - 1},
\end{equation}
where $\tilde{h}(z) = H(z)/H_{0}$, and with $Om = \Omega^{(0)}_\mathrm{dm}$ being a null test for the $\Lambda$CDM model. Moreover, it turns out that $H(z)$ is a model independent quantity and that $Om$ for the FRW universe is a combination of the Hubble rate $H(z)$ and the redshift, which makes it very attractive. The $Om$ analysis has been used quite intensively in the recent literature, in particular, to distinguish several dark energy models from $\Lambda$CDM. More specifically, $Om > \Omega^{(0)}_\mathrm{dm}$ and $Om < \Omega^{(0)}_\mathrm{dm}$ stand for the quintessence and phantom nature of the model, respectively. The definition of $Om$ shows that it is not absolutely necessary to know  the precise constraints on the matter density parameter, as well as $H_{0}$ (i.e. the value of the Hubble parameter at $z=0$),  in order to determine the nature of the model studied.

\begin{figure}[h!]
 \begin{center}$
 \begin{array}{cccc}
\includegraphics[width=85 mm]{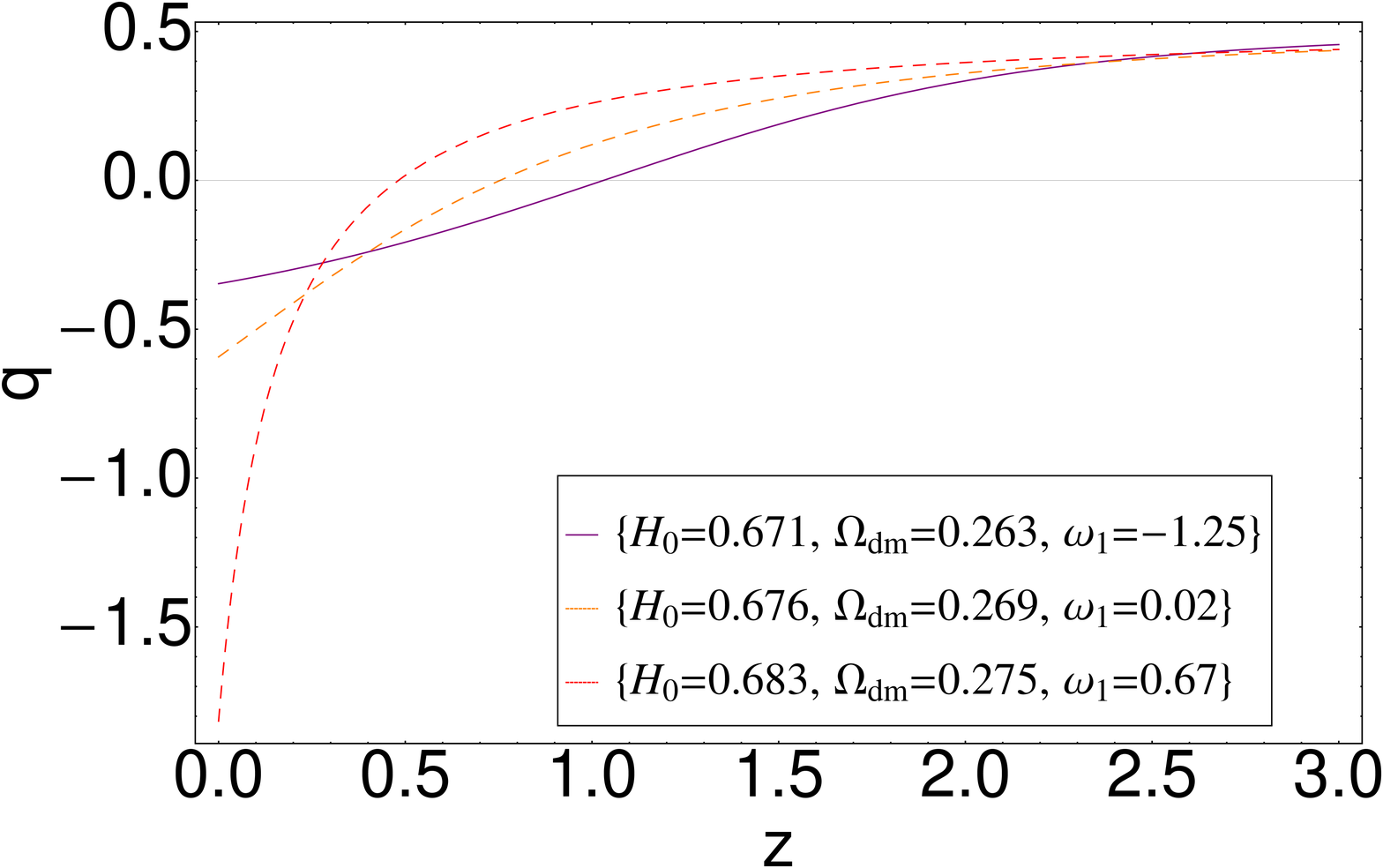} &
\includegraphics[width=85 mm]{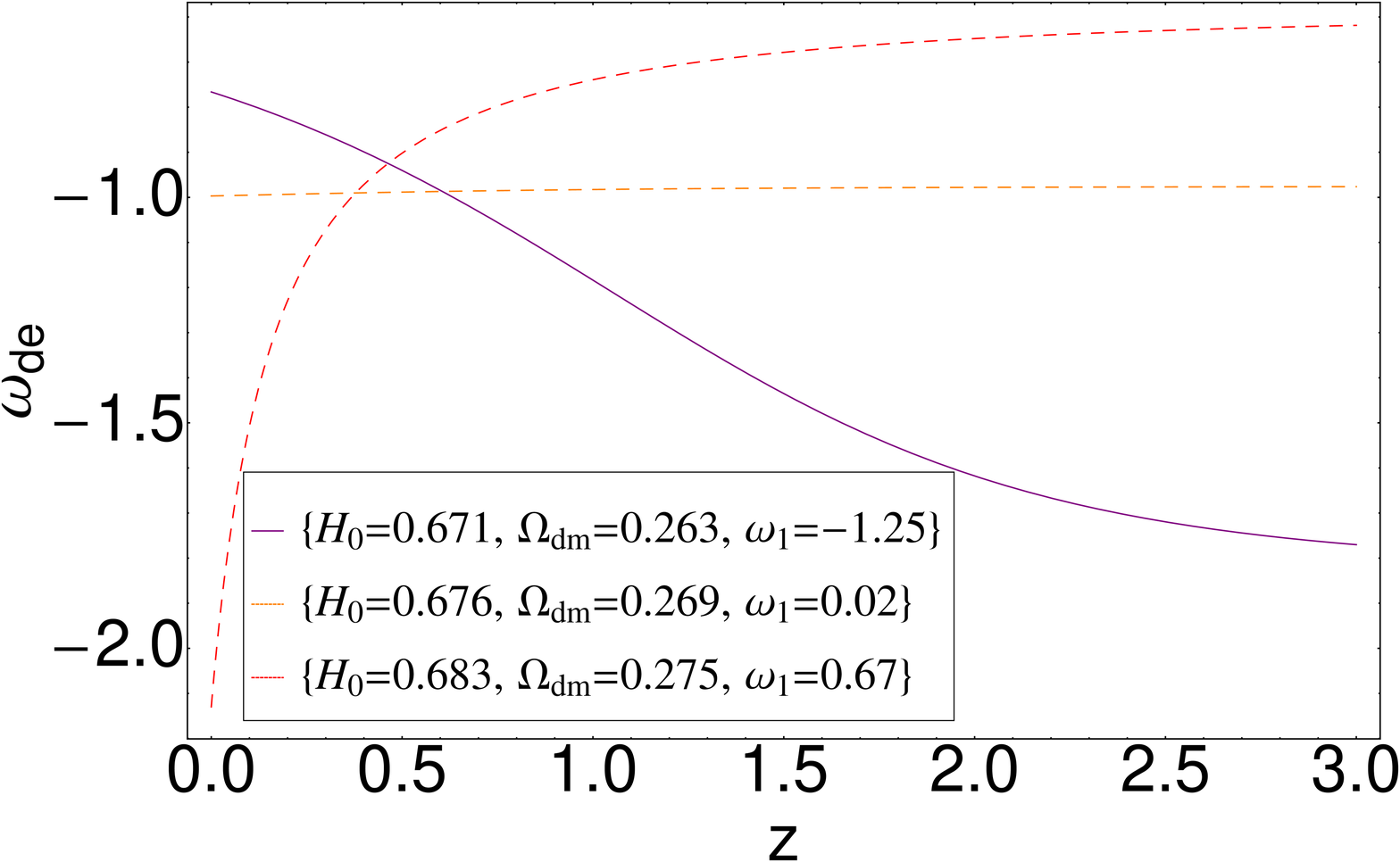}\\
 \end{array}$
 \end{center}
\caption{Plot of the behaviors of the deceleration parameter $q$ and of the EoS parameter of the dark fluid described by Eq.~(\ref{eq:omegaDE}). The purple, orange and red curves represent the behavior for $\omega_{0} = -1.2, -0.985$, and $=0.913$, respectively.}
 \label{fig:1}
\end{figure}

A plot of the behaviors of the deceleration parameter $q$ and of $\omega_\mathrm{de}$ is presented in Fig.~\ref{fig:1}. We see that the simple analysis performed here already predicts the transition redshift to be $z_\mathrm{tr} \in (0.5,1.0)$. The graphical behavior of $\omega_\mathrm{de}$ shows possible differences between the $\Lambda$CDM standard model and the new model here considered. In particular, we see that in case of  $H_{0} = 0.676$, $ \Omega_\mathrm{dm}= 0.269$, $\omega_{0}=-0.985$ and $\omega_{1} = 0.02$, the difference is negligible to high precision.  On the other hand, we observed also that if we choose the values of the parameters in a such a way that the model explains the statistically allowed lower value of the Hubble parameter at $z=2.34$, then the behavior of $\omega_\mathrm{de}$ indicates a phantom nature of the dark fluid at the early universe, with a smooth transition to  quintessence at the low-redshift universe. However, a Type I or Type III singularity~(with the phantom line crossing at $z\approx 0.38$) will be observed at $z=0$ for the chosen values of the parameters  of the model explaining the statistically allowed upper value for the Hubble parameter as reported from the BOSS experiment~(1$\sigma$ error). In this case, the dark fluid will have quintessence nature at the early universe.

\begin{figure}[h!]
 \begin{center}$
 \begin{array}{cccc}
\includegraphics[width=85 mm]{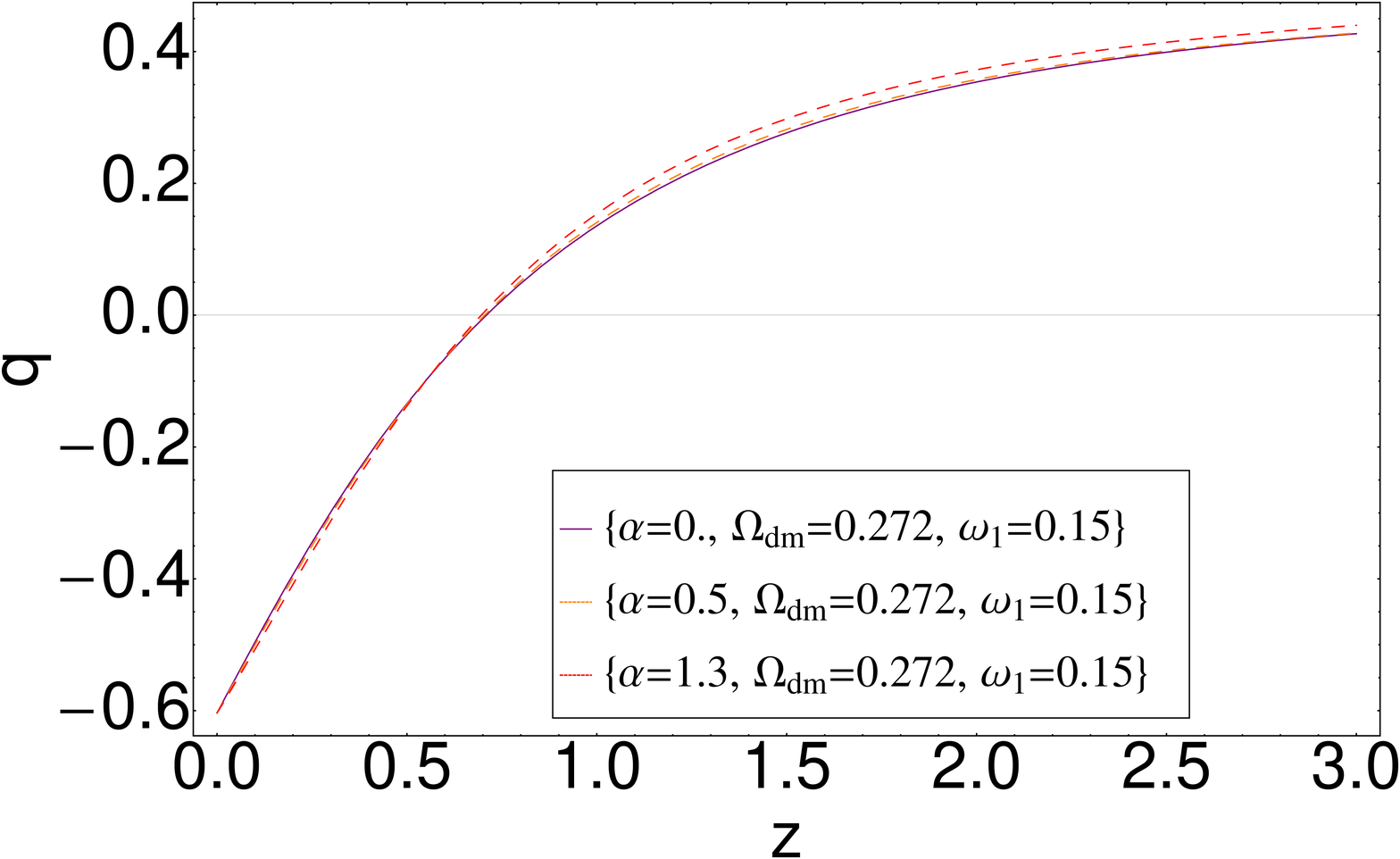} &
\includegraphics[width=85 mm]{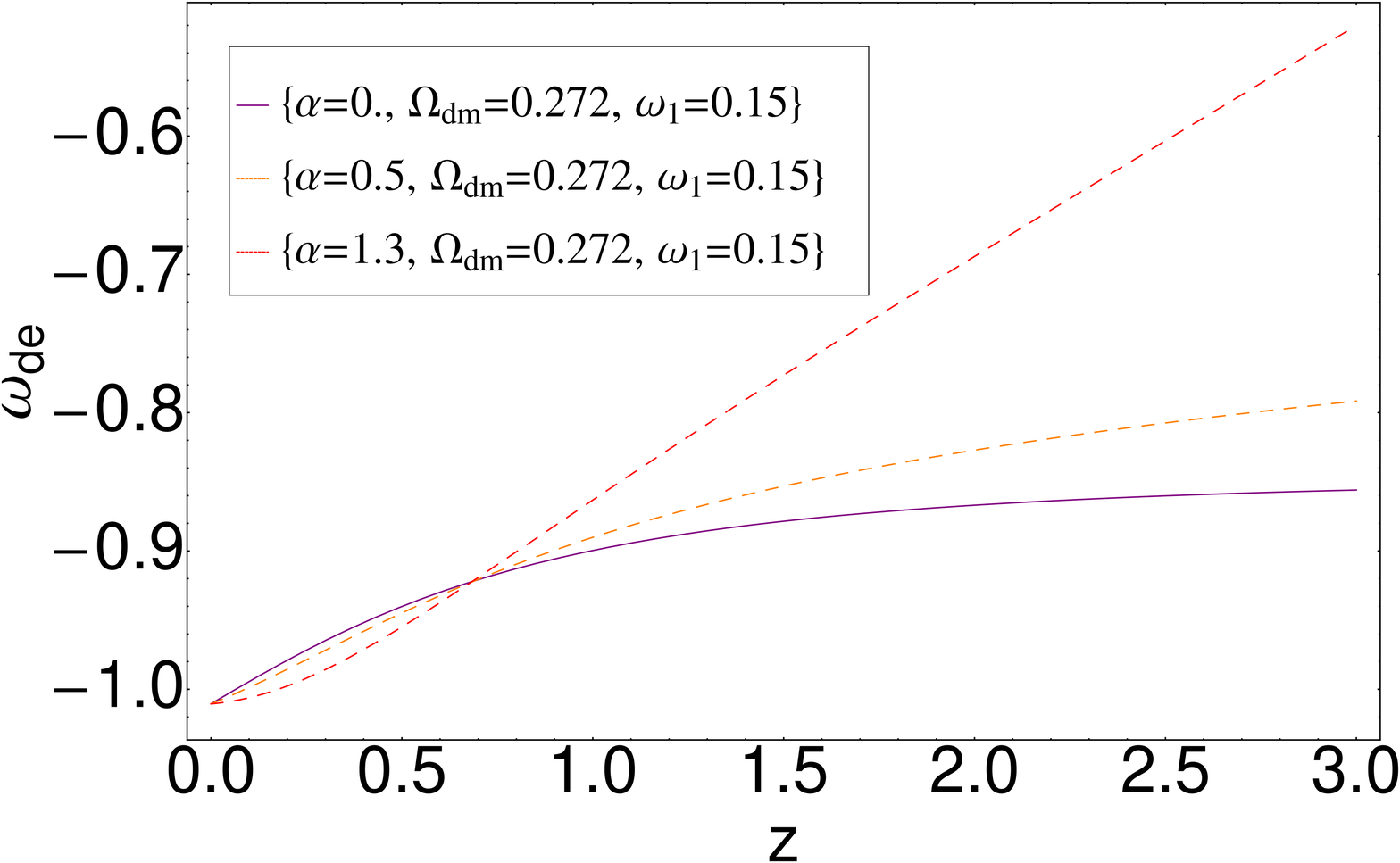}\\
 \end{array}$
 \end{center}
\caption{Graphical behavior of the deceleration parameter $q$ and the EoS parameter for the dark fluid described by Eq.~(\ref{eq:omegaDEGen}), when $\omega_{0} = -0.92$.}
 \label{fig:2}
\end{figure}

The  behavior of the deceleration and of the EoS parameters shown in Fig.~\ref{fig:2} corresponds to the model given by Eq.~(\ref{eq:omegaDEGen}) for the specific values of the parameters satisfying constraints from the $Omh^{2}$ analysis. In particular, the study of $\omega_\mathrm{de}$, Eq.~(\ref{eq:omegaDEGen}), points out the quintessential nature of the dark fluid. However, a phantom line crossing at low redshifts~(the value of $\omega_\mathrm{de}$ satisfying the constraints imposed by the PLANCK 2015 survey) is also possible. On the other hand, we see that a growth of the $\alpha$ parameter will be reflected as an increase of the value of $\omega_\mathrm{de}$ for higher redshifts. To the contrary, the impact of the parameter $\alpha$ on the transition redshift and on the present day value of the deceleration parameter is negligible.

In general, a detailed study of the models~(a part of which is presented above demonstrating the behavior of the parameters for certain values of parameters) indicates a deep necessity to work on a classification of future singularities. In the light of the above discussion it is convenient to study this point directly using numerical solutions instead of assuming a specific parametrization of the scale factor~(or the Hubble parameter).  The particular models of non-gravitational interactions considered are as follows: $Q = 3 H b \rho_\mathrm{de}$, $Q = 3 H b \rho_\mathrm{dm}$ and $Q = 3 H b (\rho_\mathrm{de} + \rho_\mathrm{dm})$.

In particular, the study of the first model, corresponding to Eq.~(\ref{eq:omegaDE}), showed that
\begin{itemize}
\item When the values of the parameters permit to obtain the statistically allowed~(with 1$\sigma$ error) highest value of the Hubble parameter at $z=2.34$, then the appearance of of Type I or Type III singularities is possible already at $z=0$. Moreover, the forms for the interaction here considered  cannot prevent the formation of the singularity or change its type.
\item On the other hand, for instance, for $b \in [-0.1 -0.01)$ a Type III singularity will be observed, while for $b \in [-0.01, 0.1]$ a Type I singularity will appear. This is due to the fact that interactions considered here do not affect the present-time behavior of the EoS parameter of dark and/or effective fluids~(for the constraints on the parameter $b$ considered here),  $b < 0$ indicating dark matter transition to dark energy. Therefore, the formation of the singularity, which already starts at $z \approx 0$ due to the phantom nature of the dark and effective fluids, will not be further affected.
\item The only possible effect of the energy  transfer will be a change of  the type of singularity.
\end{itemize}

Moreover, for $b > 0$ in the cases $Q = 3Hb\rho_\mathrm{de}$ and $Q = 3Hb\rho_\mathrm{dm}$ a singularity formation in the future will not be observed, because $\omega_\mathrm{de}$ and $\omega_\mathrm{eff}$ are quintessence fluids. The same is true for the model with $Q = 3Hb (\rho_\mathrm{de} + \rho_\mathrm{dm})$ and $b < 0.3$. However, for $b > 0.3$ the background dynamics is such that Type V singularity formation is possible, while  energy transfer from dark matter to dark energy~($b<0$) will lead to the formation of a Type I singularity.

Now, if we take the values of the parameters which permit to  obtain the lowest statistically allowed value for the Hubble parameter at $z=2.34$, we find in this case that:
\begin{itemize}
\item For $Q = 3Hb\rho_\mathrm{de}$ and $b>0$ (or $b < 0$) singularity formation is not possible, since independently of the nature of dark energy, the effective fluid has quintessence nature.
\item The interaction $Q = 3Hb\rho_\mathrm{dm}$ does not lead the formation of any kind of singularity.
\item There are cases however, as for instance,  $b > 0.2$ with $Q = 3 H b (\rho_\mathrm{de} + \rho_\mathrm{dm})$, where an $\omega$-singularity will be formed. For $b < 0$, on the contrary, future singularities will not be generated.
\end{itemize}

The second type of EoS parameter, Eq.~(\ref{eq:omegaDEGen}), gives the following picture:
\begin{itemize}
\item For the interaction $Q = 3Hb(\rho_\mathrm{de} + \rho_\mathrm{dm})$, if $b > 0.07$ an $\omega$-singularity will be formed, while if $b < -0.08$ a Type I singularity will be generated. On the other hand, if $b \in (-0.08,0.07)$, due to the fact that $\omega_\mathrm{tot} > -1$ and $\omega > -1$, no singularity will be formed. For this case $\alpha = 0.5$
\item In case when $\alpha = 1.3$, again for $b < -0.1$ a Type I singularity will appear, while for $b > 0.2$ an $\omega$-singularity will be generated, always provided the interaction be, as before, $Q = 3Hb(\rho_\mathrm{de} + \rho_\mathrm{dm})$.

\end{itemize}

Summing up, the analysis of the solutions obtained directly from the field equations indicates that the parameter space can be divided into several regions providing a very rich and interesting scenario concerning  the future evolution of the Universe. It should be clear that the used $Omh^{2}$ analysis with $3$-point values does not provide necessary information for the final conclusion, as already mentioned, and it is extremely important to involve other observational data sets. In general, it is extremely important to do a $\chi^{2}$ test and obtain the constraints, and only after that one can draw a reliable conclusion about the future evolution of the Universe according to the considered EoS parametrization. We expect to report about the extended analysis of the models considered here in a future paper. Moreover, we have found that the parameter $\alpha$ does not affect the type of singularity that can be formed owing to the background dynamics, and this is true even in the presence of non-gravitational interactions of the forms considered here. It is an interesting fact and should be mentioned that during the reconstruction of the non-gravitational interaction for the model with, Eq.~(\ref{eq:omegaDEGen}), from $H(z)$ 40-point sample data, we observed that the parameter $\alpha$ does not affect the present  behavior of the non-gravitational interaction, i.e. a corresponding  change of the type of the singularity~(or of the formation of the singularity).

\section{Discussion and conclusions}\label{sec:D}

In this paper we have introduced a new formulation of the EoS parameter for the dark fluid in terms of the deceleration parameter of the cosmological theory. In some of the cases considered, a careful phase space analysis was carried out, with indication of the corresponding critical points. For some specific values of the parameters, we produced several phase space plots indicating the type of the attractor (for stable critical points) in each  case. Our study has revealed that, in general, we can divide the parameter space into several regions, what allowed us to stabilize the critical points. In particular, in the case of non-interacting models, we have observed that the critical points are  stable nodes, while for interacting models of the first type a critical point can also be a stable focus (always depending on the form of the non-gravitational interaction considered, and on the prior constraints). An example of such stable focus is the C.P.1 critical point.

In addition, we have employed Gaussian process techniques to reconstruct the behavior of the Hubble, deceleration and $\Omega_\mathrm{de}$ parameters, using the deceleration parameter in order to reconstruct the EoS of dark and effective fluids. Gaussian processes constitute indeed a powerful tool, allowing to reconstruct the behavior of a function (and its derivatives) directly from the given data of the corresponding kernel. They have been used intensively to study various cosmological models. Results from recent literature show, for instance, that they can be applied successfully to reconstruct the behavior of a possible non-gravitational interaction between dark energy and dark matter (among other results). The above mentioned quantities have been reconstructed from a sample of $40$ $H(z)$ values. From them, $30$ out of the $40$ values were deduced with the differential age method, while the remaining $10$ are values obtained from the radial BAO method.

GP reconstruction of the deceleration parameter immediately shows that one should expect that, at higher redshifts, the Universe will be in an accelerated phase. Moreover, that another transition should occur, at an intermediate redshift, from the mentioned accelerated phase to a decelerated one. Finally, that at lower redshifts we will again recover the observed accelerated expansion of the large scale Universe. In order to continue the study of the models, we have taken $\Omega_\mathrm{de}$ as control parameter, i.e., in parallel to the EoS for dark energy, we investigated the behavior of $\Omega_\mathrm{de}$  and derived constraints on the parameters of the model. Eventually, after several runs we were able indeed to obtain constraints on the parameters.

In summary, based on GP reconstruction from $H(z)$ data, we have finally concluded that the model of Eq.~(\ref{eq:omegaDE}), according to its $68\%$ CL upon being reconstructed from the sample of $40$ $H(z)$ values, can be accepted. In addition, the same model with $68\%$ CL obtained from the sample of $30$  $H(z)$ values could be accepted, too. And also the model, Eq.~(\ref{eq:omegaDEGen}), under exactly the same conditions. It should be mentioned that due to the specific behavior of the deceleration parameter observed during reconstruction from $H(z)$ data, $\omega_\mathrm{0}$, $\omega_\mathrm{1}$, and $\alpha$ are such that only $\omega_\mathrm{de} < 0$ will be possible, for both models. Additionally, a very interesting model, on top of the two models considered, is the one described by Eq.~(\ref{eq:omegaDE}), since it can reach to explain the value of the Hubble parameter at the higher redshift, $z=2.34$, as reported by the BOSS experiment. However, in order to understand this properly with the reconstructed interaction, we need in this case to involve a non-gravitational interaction between dark energy and dark matter, to explain the mentioned value of the Hubble parameter. However, even if it would seem we do need to include a non-gravitational interaction, this occurs for redshifts not covered by recent $H(z)$ data. Therefore, at this moment we do not pretend to have a final answer to this question. On the other hand, even if,  in principle, we were able to reconstruct the Hubble parameter beyond $z=2.6$, the errors arising from the GP reconstruction are large in this case, again  not allowing to reach a clear conclusion. As a consequence, it would be important to extend the present study to involve additional observational data sets~(in particular GRB data).  It will be interesting to consider how our conclusion may change when the volume of the available data grows. In this case, however, we will have to start the reconstruction necessarily from the second order differential equation describing the growth rate dynamics. 

GP reconstruction cannot actually reconstruct information concerning the possible singularities that could be formed in certain cosmological models, because it is based on infinitely differentiable covariance functions. Therefore, in order to extend this study further we need analyze numerical solutions obtained directly from the field equations. We would like to mention also, that  for appropriate initial conditions and values of the model parameters, we have studied  the resulting model and found that the value of the Hubble parameter, as reported from the BOSS experiment, could be easily recovered. It should be duly remarked that, just with the help of our parametrization, this goal could be accomplished quite naturally, and that there was no need to introduce any sort of interaction between the energy components whatsoever. To wit, in the recent literature it has been contrarily defended that, in order to explain the results of the BOSS experiment, a non-gravitational interaction needed be used, or either one had to resort to a dark energy model with an awkward  EoS. We must here point out, in this respect, that the EoS for the dark fluids we have considered in this paper are in no way strange, our approach to the problem having been completely different.

This is a very valuable result for future analysis, since, for instance, from the GP reconstruction we observed that the parameter $\alpha$ does not affect the reconstruction of the non-gravitational interaction. What means, on its turn, that it cannot affect the singularity formation, and neither the type of singularity. On the other hand, we found a similar conclusion during the study of numerical solutions to the field equations. Therefore, there is the hope that GP reconstruction with additional data sets will additionally show that the models here can really explain the value of the Hubble parameter reported by the BOSS experiment without accounting for non-gravitational interactions.  We have specified several values for the parameters, in order to demonstrate that the second model, described by Eq.~(\ref{eq:omegaDEGen}), is in good agreement with the results coming from the BOSS experiment. Actually, both models can explain the accelerated expansion of the Universe and are free from the cosmological coincidence problem, which makes them both viable, being also, in a way, quite simple.

In order to gain a better understanding of the models, we have studied the formation of possible future singularities, for three different forms of the non-gravitational interaction. Our study allows to identify the type of singularity in each case and to see how the interaction considered will affect the formation of the singularity, in the models where the effective fluid has quintessence nature. In particular, we have found that, in the cases of the models here considered, this will depend on the form of the non-gravitational interaction and on the direction of the energy transfer from one energy type to the other. According to that, either a Type I, Type III, or an $\omega$-singularity~(Type V) can be generated.
As a result, we have compiled very valuable additional information which might be useful in the future, under the form of templates, to confront new observational data and, in this way, finally adjust the model that better fits the results of forthcoming astronomical surveys.

The influence of the results obtained in this paper on the cosmological structure formation process and the investigation of more general forms for the non-gravitational interactions, including the interesting case of sign-changing interactions --to identify possible additional singularities in the cosmological models considered-- will be reported elsewhere. Finally, a further extension of this work will be the construction of some appropriate modified theories of gravity incorporating the transitions discussed here, together with their astrophysical applications.

\medskip

\section*{Acknowledgements}

This work has been partially supported by CSIC, I-LINK1019 Project (EE and SN), by MINECO (Spain), Projects FIS2013-44881-P and FIS2016-76363-P, by the CPAN Consolider Ingenio 2010 Project, and by JSPS (Japan), Fellowship N. S17017 (EE), and also by a MEXT KAKENHI Grant-in-Aid for Scientific Research on Innovative Areas ``Cosmic Acceleration'' (No. 15H05890) (SN). MK is supported in part by the Chinese Academy of Sciences President's International Fellowship Initiative, Grant No. 2018PM0054.
This research was started while EE was visiting the Kobayashi-Maskawa Institute in Nagoya, Japan (EE is much obliged with SN and the rest of the members of the KMI for very kind hospitality), and was finished when SN visited the Institute for Space Sciences, ICE-CSIC in Bellaterra, Spain (SN thanks EE, and the rest of the members of the ICE for very kind hospitality).


\begin{thebibliography}{99}

%\cite{Xia:2016vnp}
\bibitem{Xia:2016vnp}
D.~M.~Xia and S.~Wang,
%``Constraining interacting dark energy models with latest cosmological observations,''
Mon.\ Not.\ Roy.\ Astron.\ Soc.\  {\bf 463} (2016) no.1,  952
doi:10.1093/mnras/stw2073
[arXiv:1608.04545 [astro-ph.CO]].
%%CITATION = doi:10.1093/mnras/stw2073;%%
%10 citations counted in INSPIRE as of 10 Nov 2017

%\cite{Zhang:2013zyn}
\bibitem{Zhang:2013zyn}
M.~J.~Zhang and W.~B.~Liu,
%``Observational constraint on the interacting dark energy models including the Sandage-Loeb test,''
Eur.\ Phys.\ J.\ C {\bf 74} (2014) 2863
doi:10.1140/epjc/s10052-014-2863-x
[arXiv:1312.0224 [astro-ph.CO]].
%%CITATION = doi:10.1140/epjc/s10052-014-2863-x;%%
%14 citations counted in INSPIRE as of 10 Nov 2017

%\cite{Yang:2017iew}
\bibitem{Yang:2017iew}
W.~Yang, L.~Xu, H.~Li, Y.~Wu and J.~Lu,
%``Testing the Interacting Dark Energy Model with Cosmic Microwave Background Anisotropy and Observational Hubble Data,''
Entropy {\bf 19} (2017) no.7,  327.
doi:10.3390/e19070327
%%CITATION = doi:10.3390/e19070327;%%

%\cite{Marsh:2016ynw}
\bibitem{Marsh:2016ynw}
M.~C.~D.~Marsh,
%``Exacerbating the cosmological constant problem with interacting dark energy models,''
Phys.\ Rev.\ Lett.\  {\bf 118} (2017) no.1,  011302
doi:10.1103/PhysRevLett.118.011302
[arXiv:1606.01538 [astro-ph.CO]].
%%CITATION = doi:10.1103/PhysRevLett.118.011302;%%
%6 citations counted in INSPIRE as of 10 Nov 2017

%\cite{Sadjadi:2006qp}
\bibitem{Sadjadi:2006qp}
H.~M.~Sadjadi and M.~Alimohammadi,
%``Cosmological coincidence problem in interactive dark energy models,''
Phys.\ Rev.\ D {\bf 74} (2006) 103007
doi:10.1103/PhysRevD.74.103007
[gr-qc/0610080].
%%CITATION = doi:10.1103/PhysRevD.74.103007;%%
%96 citations counted in INSPIRE as of 11 Nov 2017

%\cite{Hakobyan:2013hca}
\bibitem{Hakobyan:2013hca}
J.~Sadeghi, M.~Khurshudyan, J.~Hakobyan and H.~Farahani,
%``Mutually interacting Tachyon dark energy with variable $G$ and $\Lambda$,''
Res.\ Astron.\ Astrophys.\  {\bf 15} (2015) no.2,  175
doi:10.1088/1674-4527/15/2/002
[arXiv:1309.7496 [gr-qc]].
%%CITATION = doi:10.1088/1674-4527/15/2/002;%%
%13 citations counted in INSPIRE as of 11 Nov 2017

%\cite{Sadeghi:2015nda}
\bibitem{Sadeghi:2015nda}
J.~Sadeghi, B.~Pourhassan, A.~S.~Kubeka and M.~Rostami,
%``Logarithmic corrected Polynomial $f(R)$ inflation mimicking a cosmological constant,''
Int.\ J.\ Mod.\ Phys.\ D {\bf 25} (2016) no.07,  1650077
doi:10.1142/S0218271816500772
[arXiv:1510.02351 [gr-qc]].
%%CITATION = doi:10.1142/S0218271816500772;%%
%2 citations counted in INSPIRE as of 11 Nov 2017

%\cite{Khurshudyan:2013oba}
\bibitem{Khurshudyan:2013oba}
M.~Khurshudyan, A.~Khurshudyan and R.~Myrzakulov,
%``Interacting varying Ghost Dark energy models in General Relativity,''
Astrophys.\ Space Sci.\  {\bf 357} (2015) no.2,  113
doi:10.1007/s10509-015-2341-4
[arXiv:1307.7859 [gr-qc]].
%%CITATION = doi:10.1007/s10509-015-2341-4;%%
%26 citations counted in INSPIRE as of 11 Nov 2017

%\cite{Khurshudyan:2015mva}
\bibitem{Khurshudyan:2015mva}
M.~Khurshudyan and R.~Myrzakulov,
%``Phase space analysis of some interacting Chaplygin gas models,''
Eur.\ Phys.\ J.\ C {\bf 77} (2017) no.2,  65
doi:10.1140/epjc/s10052-017-4634-y
[arXiv:1509.02263 [gr-qc]].
%%CITATION = doi:10.1140/epjc/s10052-017-4634-y;%%
%22 citations counted in INSPIRE as of 11 Nov 2017

%\cite{Khurshudyan:2015mpa}
\bibitem{Khurshudyan:2015mpa}
M.~Khurshudyan,
%``Some non linear interactions in polytropic gas cosmology: Phase space analysis,''
Astrophys.\ Space Sci.\  {\bf 360} (2015) no.1,  33
doi:10.1007/s10509-015-2540-z
[arXiv:1510.07962 [physics.gen-ph]].
%%CITATION = doi:10.1007/s10509-015-2540-z;%%
%11 citations counted in INSPIRE as of 11 Nov 2017

%\cite{Khurshudyan:2016uql}
\bibitem{Khurshudyan:2016uql}
M.~Khurshudyan,
%``On a holographic dark energy model with a Nojiri-Odintsov cut-off in general relativity,''
Astrophys.\ Space Sci.\  {\bf 361} (2016) no.7,  232
doi:10.1007/s10509-016-2821-1
[arXiv:1606.05264 [gr-qc]].
%%CITATION = doi:10.1007/s10509-016-2821-1;%%
%8 citations counted in INSPIRE as of 11 Nov 2017

%\cite{Khurshudyan:2016dki}
\bibitem{Khurshudyan:2016dki}
M.~Z.~Khurshudyan and A.~N.~Makarenko,
%``On a phenomenology of the accelerated expansion with a varying ghost dark energy,''
Astrophys.\ Space Sci.\  {\bf 361} (2016) no.6,  187
doi:10.1007/s10509-016-2775-3
[arXiv:1606.06590 [gr-qc]].
%%CITATION = doi:10.1007/s10509-016-2775-3;%%
%6 citations counted in INSPIRE as of 11 Nov 2017

%\cite{Khurshudyan:2016ziq}
\bibitem{Khurshudyan:2016ziq}
M.~Khurshudyan,
%``Varying ghost dark energy and particle creation,''
Eur.\ Phys.\ J.\ Plus {\bf 131} (2016) no.2,  25.
doi:10.1140/epjp/i2016-16025-7
%%CITATION = doi:10.1140/epjp/i2016-16025-7;%%
%10 citations counted in INSPIRE as of 11 Nov 2017

%\cite{Khurshudyan:2016epw}
\bibitem{Khurshudyan:2016epw}
M.~Khurshudyan,
%``Thermodynamics of ghost dark energy in case of various nonlinear interactions,''
Eur.\ Phys.\ J.\ Plus {\bf 131} (2016) no.5,  174.
doi:10.1140/epjp/i2016-16174-7
%%CITATION = doi:10.1140/epjp/i2016-16174-7;%%

%\cite{Khurshudyan:2016atd}
\bibitem{Khurshudyan:2016atd}
M.~Khurshudyan,
%``Low redshift universe and a varying ghost dark energy,''
Mod.\ Phys.\ Lett.\ A {\bf 31} (2016) no.09,  1650055.
doi:10.1142/S0217732316500553
%%CITATION = doi:10.1142/S0217732316500553;%%
%9 citations counted in INSPIRE as of 11 Nov 2017

%\cite{Khurshudyan:2016xst}
\bibitem{Khurshudyan:2016xst}
M.~Khurshudyan,
%``A varying polytropic gas universe and phase space analysis,''
Mod.\ Phys.\ Lett.\ A {\bf 31} (2016) no.16,  1650097.
doi:10.1142/S0217732316500978
%%CITATION = doi:10.1142/S0217732316500978;%%
%7 citations counted in INSPIRE as of 11 Nov 2017

%\cite{Khurshudyan:2016zse}
\bibitem{Khurshudyan:2016zse}
M.~Khurshudyan,
%``On the Phenomenology of an Accelerated Large-Scale Universe,''
Symmetry {\bf 8} (2016) no.12,  110.
doi:10.3390/sym8110110
%%CITATION = doi:10.3390/sym8110110;%%
%6 citations counted in INSPIRE as of 11 Nov 2017

%\cite{Khurshudyan:2017kmf}
\bibitem{Khurshudyan:2017kmf}
M.~Khurshudyan and A.~Khurshudyan,
%``Some interacting dark energy models,''
arXiv:1708.02293 [gr-qc].
%%CITATION = ARXIV:1708.02293;%%
%1 citations counted in INSPIRE as of 11 Nov 2017

%\cite{Khurshudyan:2017qtd}
\bibitem{Khurshudyan:2017qtd}
M.~Khurshudyan and A.~Khurshudyan,
%``On cosmology of interacting varying polytropic dark fluids,''
arXiv:1707.04116 [gr-qc].
%%CITATION = ARXIV:1707.04116;%%

%\cite{Brevik:2017msy}
\bibitem{Brevik:2017msy}
I.~Brevik, {\O}.~Gr{\o}n, J.~de Haro, S.~D.~Odintsov and E.~N.~Saridakis,
%``Viscous Cosmology for Early- and Late-Time Universe,''
Int.\ J.\ Mod.\ Phys.\ D {\bf 26} (2017) 1730024
doi:10.1142/S0218271817300245
[arXiv:1706.02543 [gr-qc]].
%%CITATION = doi:10.1142/S0218271817300245;%%
%15 citations counted in INSPIRE as of 11 Nov 2017

%\cite{Elizalde:2017mrn}
\bibitem{Elizalde:2017mrn}
E.~Elizalde, S.~D.~Odintsov, L.~Sebastiani and R.~Myrzakulov,
%``Beyond-one-loop quantum gravity action yielding both inflation and late-time acceleration,''
Nucl.\ Phys.\ B {\bf 921} (2017) 411
doi:10.1016/j.nuclphysb.2017.06.003
[arXiv:1706.01879 [gr-qc]].
%%CITATION = doi:10.1016/j.nuclphysb.2017.06.003;%%
%4 citations counted in INSPIRE as of 11 Nov 2017

%\cite{Vasilis:2018ov1}
\bibitem{Vasilis:2018ov1}
S. ~D.~Odintsov, V.~ K.~Oikonomou,
%``Dynamical systems perspective of cosmological finite-time singularities in f(R) gravity and interacting multifluid cosmology,''
Phys.\ Rev.\ D {\bf 98} (2018) no.2, 024013
doi:10.1103/PhysRevD.98.024013
[arXiv:1806.07295 [gr-qc]].
%%CITATION = doi:10.1103/PhysRevD.98.024013;%%
%0 citations counted in INSPIRE as of 09 August 2018

%\cite{Vasilis:2018ov2}
\bibitem{Vasilis:2018ov2}
S. ~D.~Odintsov, V.~ K.~Oikonomou,
%``A Study of Finite-time Singularities of Loop Quantum Cosmology Interacting Multifluids,''
Phys.\ Rev.\ D {\bf 97} (2018), 124042
doi:10.1103/PhysRevD.97.124042
[arXiv:1806.01588 [gr-qc]].
%%CITATION = doi:10.1103/PhysRevD.97.124042;%%
%0 citations counted in INSPIRE as of 09 August 2018

%\cite{Nojiri:2005pu}
\bibitem{Nojiri:2005pu}
S.~Nojiri and S.~D.~Odintsov,
%``Unifying phantom inflation with late-time acceleration: Scalar phantom-non-phantom transition model and generalized holographic dark energy,''
Gen.\ Rel.\ Grav.\  {\bf 38} (2006) 1285
doi:10.1007/s10714-006-0301-6
[hep-th/0506212].
%%CITATION = doi:10.1007/s10714-006-0301-6;%%
%472 citations counted in INSPIRE as of 11 Nov 2017

%\cite{Nojiri:2017opc}
\bibitem{Nojiri:2017opc}
S.~Nojiri and S.~D.~Odintsov,
%``Covariant Generalized Holographic Dark Energy and Accelerating Universe,''
Eur.\ Phys.\ J.\ C {\bf 77} (2017) no.8,  528
doi:10.1140/epjc/s10052-017-5097-x
[arXiv:1703.06372 [hep-th]].
%%CITATION = doi:10.1140/epjc/s10052-017-5097-x;%%
%6 citations counted in INSPIRE as of 11 Nov 2017

%\cite{Elizalde:2016tmn}
\bibitem{Elizalde:2016tmn}
E.~Elizalde and L.~G.~T.~Silva,
%``Inhomogeneous imperfect fluid inflation,''
Astrophys.\ Space Sci.\  {\bf 362} (2017) no.1,  7
doi:10.1007/s10509-016-2988-5
[arXiv:1611.08255 [gr-qc]].
%%CITATION = doi:10.1007/s10509-016-2988-5;%%
%1 citations counted in INSPIRE as of 11 Nov 2017

%\cite{Brevik:2016kwq}
\bibitem{Brevik:2016kwq}
I.~Brevik, E.~Elizalde, V.~V.~Obukhov and A.~V.~Timoshkin,
%``Inflationary Universe with a Viscous Fluid Avoiding Self-Reproduction,''
Annalen Phys.\  {\bf 529} (2017) 0195
doi:10.1002/andp.201600195
[arXiv:1609.04953 [gr-qc]].
%%CITATION = doi:10.1002/andp.201600195;%%
%2 citations counted in INSPIRE as of 11 Nov 2017

%\cite{Odintsov:2017icc}
\bibitem{Odintsov:2017icc}
S.~D.~Odintsov, V.~K.~Oikonomou and P.~V.~Tretyakov,
%``Phase space analysis of the accelerating multifluid Universe,''
Phys.\ Rev.\ D {\bf 96} (2017) no.4,  044022
doi:10.1103/PhysRevD.96.044022
[arXiv:1707.08661 [gr-qc]].
%%CITATION = doi:10.1103/PhysRevD.96.044022;%%
%2 citations counted in INSPIRE as of 11 Nov 2017

%\cite{Bamba:2015uma}
\bibitem{Bamba:2015uma}
K.~Bamba and S.~D.~Odintsov,
%``Inflationary cosmology in modified gravity theories,''
Symmetry {\bf 7} (2015) no.1,  220
doi:10.3390/sym7010220
[arXiv:1503.00442 [hep-th]].
%%CITATION = doi:10.3390/sym7010220;%%
%109 citations counted in INSPIRE as of 11 Nov 2017


%\cite{Kremer:2003vs}
\bibitem{Kremer:2003vs}
G.~M.~Kremer,
%``Cosmological models described by a mixture of van der Waals fluid and dark energy,''
Phys.\ Rev.\ D {\bf 68} (2003) 123507
doi:10.1103/PhysRevD.68.123507
[gr-qc/0309111].
%%CITATION = doi:10.1103/PhysRevD.68.123507;%%
%57 citations counted in INSPIRE as of 11 Nov 2017

%\cite{Brevik:2017juz}
\bibitem{Brevik:2017juz}
I.~Brevik, E.~Elizalde, S.~D.~Odintsov and A.~V.~Timoshkin,
%``Inflationary universe in terms of a van der Waals viscous fluid,''
Int. J. Geom. Meth. Mod. Phys. {\bf 14}, 1750185 (2017), doi:10.1142/S0219887817501857,
arXiv:1708.06244 [gr-qc].
%%CITATION = ARXIV:1708.06244;%%
%1 citations counted in INSPIRE as of 11 Nov 2017

%\cite{Capozziello:2005pa}
\bibitem{Capozziello:2005pa}
S.~Capozziello, V.~F.~Cardone, E.~Elizalde, S.~Nojiri and S.~D.~Odintsov,
%``Observational constraints on dark energy with generalized equations of state,''
Phys.\ Rev.\ D {\bf 73}, 043512 (2006)
doi:10.1103/PhysRevD.73.043512
[astro-ph/0508350].
%%CITATION = doi:10.1103/PhysRevD.73.043512;%%
%230 citations counted in INSPIRE as of 13 Nov 2017

%\cite{Capozziello:2004ej}
\bibitem{Capozziello:2004ej}
S.~Capozziello, V.~F.~Cardone, S.~Carloni, S.~De Martino, M.~Falanga, A.~Troisi and M.~Bruni,
%``Constraining Van der Waals quintessence by observations,''
JCAP {\bf 0504} (2005) 005
doi:10.1088/1475-7516/2005/04/005
[astro-ph/0410503].
%%CITATION = doi:10.1088/1475-7516/2005/04/005;%%
%27 citations counted in INSPIRE as of 11 Nov 2017

%\cite{Gorini:2004by}
\bibitem{Gorini:2004by}
V.~Gorini, A.~Kamenshchik, U.~Moschella and V.~Pasquier,
%``The Chaplygin gas as a model for dark energy,''
doi:10.1142/9789812704030\_ 0050
gr-qc/0403062.
%%CITATION = doi:10.1142/9789812704030_0050;%%
%128 citations counted in INSPIRE as of 11 Nov 2017

%\cite{Zhang:2004gc}
\bibitem{Zhang:2004gc}
X.~Zhang, F.~Q.~Wu and J.~Zhang,
%``A New generalized Chaplygin gas as a scheme for unification of dark energy and dark matter,''
JCAP {\bf 0601} (2006) 003
doi:10.1088/1475-7516/2006/01/003
[astro-ph/0411221].
%%CITATION = doi:10.1088/1475-7516/2006/01/003;%%
%124 citations counted in INSPIRE as of 11 Nov 2017

%\cite{Khurshudyan:2014fna}
\bibitem{Khurshudyan:2014fna}
M.~Khurshudyan,
%``Interacting extended Chaplygin gas cosmology in Lyra Manifold,''
Astrophys.\ Space Sci.\  {\bf 360} (2015) no.2,  44
doi:10.1007/s10509-015-2557-3
[arXiv:1410.5858 [gr-qc]].
%%CITATION = doi:10.1007/s10509-015-2557-3;%%
%11 citations counted in INSPIRE as of 11 Nov 2017

%\cite{Copeland:2006tn}
\bibitem{Copeland:2006tn}
E.~J.~Copeland and D.~Wands,
%``Cosmological matching conditions,''
JCAP {\bf 0706} (2007) 014
doi:10.1088/1475-7516/2007/06/014
[hep-th/0609183].
%%CITATION = doi:10.1088/1475-7516/2007/06/014;%%
%18 citations counted in INSPIRE as of 11 Nov 2017

%\cite{Cai:2009zp}
\bibitem{Cai:2009zp}
Y.~F.~Cai, E.~N.~Saridakis, M.~R.~Setare and J.~Q.~Xia,
%``Quintom Cosmology: Theoretical implications and observations,''
Phys.\ Rept.\  {\bf 493} (2010) 1
doi:10.1016/j.physrep.2010.04.001
[arXiv:0909.2776 [hep-th]].
%%CITATION = doi:10.1016/j.physrep.2010.04.001;%%
%445 citations counted in INSPIRE as of 11 Nov 2017

%\cite{Guberina:2006fy}
\bibitem{Guberina:2006fy}
B.~Guberina, R.~Horvat and H.~Nikolic,
%``Dynamical dark energy with a constant vacuum energy density,''
Phys.\ Lett.\ B {\bf 636} (2006) 80
doi:10.1016/j.physletb.2006.03.041
[astro-ph/0601598].
%%CITATION = doi:10.1016/j.physletb.2006.03.041;%%
%80 citations counted in INSPIRE as of 11 Nov 2017

%\cite{Ludwick:2015dba}
\bibitem{Ludwick:2015dba}
K.~J.~Ludwick,
%``Examining the Viability of Phantom Dark Energy,''
Phys.\ Rev.\ D {\bf 92} (2015) no.6,  063019
doi:10.1103/PhysRevD.92.063019
[arXiv:1507.06492 [gr-qc]].
%%CITATION = doi:10.1103/PhysRevD.92.063019;%%
%3 citations counted in INSPIRE as of 11 Nov 2017

%\cite{Capozziello:2005tf}
\bibitem{Capozziello:2005tf}
S.~Capozziello, S.~Nojiri and S.~D.~Odintsov,
%``Unified phantom cosmology: Inflation, dark energy and dark matter under the same standard,''
Phys.\ Lett.\ B {\bf 632} (2006) 597
doi:10.1016/j.physletb.2005.11.012
[hep-th/0507182].
%%CITATION = doi:10.1016/j.physletb.2005.11.012;%%
%254 citations counted in INSPIRE as of 11 Nov 2017

%\cite{Jhingan:2008ym}
\bibitem{Jhingan:2008ym}
S.~Jhingan, S.~Nojiri, S.~D.~Odintsov, M.~Sami, I.~Thongkool and S.~Zerbini,
%``Phantom and non-phantom dark energy: The Cosmological relevance of non-locally corrected gravity,''
Phys.\ Lett.\ B {\bf 663} (2008) 424
doi:10.1016/j.physletb.2008.04.054
[arXiv:0803.2613 [hep-th]].
%%CITATION = doi:10.1016/j.physletb.2008.04.054;%%
%89 citations counted in INSPIRE as of 11 Nov 2017

%\cite{Khurshudyan:2017rrw}
\bibitem{Khurshudyan:2017rrw}
M.~Khurshudyan, M.~Hakobyan and A.~Khurshudyan,
%``Non-linear logarithmic interactions and a varying polytropic gas,''
arXiv:1707.02856 [gr-qc].
%%CITATION = ARXIV:1707.02856;%%
%1 citations counted in INSPIRE as of 11 Nov 2017

%\cite{Nojiri:2017ncd}
\bibitem{Nojiri:2017ncd}
S.~Nojiri, S.~D.~Odintsov and V.~K.~Oikonomou,
%``Modified Gravity Theories on a Nutshell: Inflation, Bounce and Late-time Evolution,''
Phys.\ Rept.\  {\bf 692} (2017) 1
doi:10.1016/j.physrep.2017.06.001
[arXiv:1705.11098 [gr-qc]].
%%CITATION = doi:10.1016/j.physrep.2017.06.001;%%
%60 citations counted in INSPIRE as of 11 Nov 2017

%\cite{Delubac:2014dt}
\bibitem{Delubac:2014dt}
T.~Delubac {\it et al.} [BOSS Collaboration],
%``Baryon acoustic oscillations in the Ly$\alpha$ forest of BOSS DR11 quasars,''
Astron.\ Astrophys.\ {\bf 574} (2015), A59
doi:10.1051/0004-6361/201423969
[arXiv:1404.1801 [astro-ph.CO]].
%%CITATION = doi:10.1051/0004-6361/201423969;%%
%322 citations counted in INSPIRE as of 19 March 2018

%\cite{Abdalla:2014cla}
\bibitem{Abdalla:2014cla}
E.~G.~M.~Ferreira, J.~Quintin, A.~A.~Costa, E.~Abdalla and B.~Wang,
%``Evidence for interacting dark energy from BOSS,``
Phys.\ Rev.\ D {\bf 95} (2017) no.4,  043520
doi:10.1103/PhysRevD.95.043520
[arXiv:1412.2777 [astro-ph.CO]].
%%CITATION = doi:10.1103/PhysRevD.95.043520;%%
%52 citations counted in INSPIRE as of 11 Nov 2017

%\cite{MSeikel:2012ms}
\bibitem{MSeikel:2012ms}
M.~Seikel, C.~Clarkson and M.~Smith,
%``Reconstruction of dark energy and expansion dynamics using Gaussian processes,``
JCAP {\bf 06} (2012) 036
doi:10.1088/1475-7516/2012/06/036
[arXiv:1204.2832 [astro-ph.CO]].
%%CITATION = doi:10.1088/1475-7516/2012/06/036 ;%%
%86 citations counted in INSPIRE as of 19 March 2018


%\cite{Nojiri:2005sx}
\bibitem{Nojiri:2005sx}
S.~Nojiri, S.~D.~Odintsov and S.~Tsujikawa,
%``Properties of singularities in (phantom) dark energy universe,''
Phys.\ Rev.\ D {\bf 71} (2005) 063004
doi:10.1103/PhysRevD.71.063004
[hep-th/0501025].
%%CITATION = doi:10.1103/PhysRevD.71.063004;%%
%835 citations counted in INSPIRE as of 11 Nov 2017

%\cite{Caldwell:1999ew}
\bibitem{Caldwell:1999ew}
R.~R.~Caldwell,
%``A Phantom menace?,''
Phys.\ Lett.\ B {\bf 545} (2002) 23
doi:10.1016/S0370-2693(02)02589-3
[astro-ph/9908168].
%%CITATION = doi:10.1016/S0370-2693(02)02589-3;%%
%2193 citations counted in INSPIRE as of 11 Nov 2017

%\cite{Caldwell:2003vq}
\bibitem{Caldwell:2003vq}
R.~R.~Caldwell, M.~Kamionkowski and N.~N.~Weinberg,
%``Phantom energy and cosmic doomsday,''
Phys.\ Rev.\ Lett.\  {\bf 91} (2003) 071301
doi:10.1103/PhysRevLett.91.071301
[astro-ph/0302506].
%%CITATION = doi:10.1103/PhysRevLett.91.071301;%%
%1450 citations counted in INSPIRE as of 11 Nov 2017

%\cite{Barrow:2004xh}
\bibitem{Barrow:2004xh}
J.~D.~Barrow,
%``Sudden future singularities,''
Class.\ Quant.\ Grav.\  {\bf 21} (2004) L79
doi:10.1088/0264-9381/21/11/L03
[gr-qc/0403084].
%%CITATION = doi:10.1088/0264-9381/21/11/L03;%%
%368 citations counted in INSPIRE as of 11 Nov 2017

%\cite{Barrow:2004hk}
\bibitem{Barrow:2004hk}
J.~D.~Barrow,
%``More general sudden singularities,''
Class.\ Quant.\ Grav.\  {\bf 21} (2004) 5619
doi:10.1088/0264-9381/21/23/020
[gr-qc/0409062].
%%CITATION = doi:10.1088/0264-9381/21/23/020;%%
%156 citations counted in INSPIRE as of 11 Nov 2017

%\cite{BouhmadiLopez:2006fu}
\bibitem{BouhmadiLopez:2006fu}
M.~Bouhmadi-Lopez, P.~F.~Gonzalez-Diaz and P.~Martin-Moruno,
%``Worse than a big rip?,''
Phys.\ Lett.\ B {\bf 659} (2008) 1
doi:10.1016/j.physletb.2007.10.079
[gr-qc/0612135].
%%CITATION = doi:10.1016/j.physletb.2007.10.079;%%
%131 citations counted in INSPIRE as of 11 Nov 2017

%\cite{Nojiri:2006ri}
\bibitem{Nojiri:2006ri}
S.~Nojiri and S.~D.~Odintsov,
%``Introduction to modified gravity and gravitational alternative for dark energy,''
eConf C {\bf 0602061} (2006) 06
 [Int.\ J.\ Geom.\ Meth.\ Mod.\ Phys.\  {\bf 4} (2007) 115]
doi:10.1142/S0219887807001928
[hep-th/0601213].
%%CITATION = doi:10.1142/S0219887807001928;%%
%1724 citations counted in INSPIRE as of 11 Nov 2017

%\cite{Nojiri:2010wj}
\bibitem{Nojiri:2010wj}
S.~Nojiri and S.~D.~Odintsov,
%``Unified cosmic history in modified gravity: from F(R) theory to Lorentz non-invariant models,''
Phys.\ Rept.\  {\bf 505} (2011) 59
doi:10.1016/j.physrep.2011.04.001
[arXiv:1011.0544 [gr-qc]].
%%CITATION = doi:10.1016/j.physrep.2011.04.001;%%
%1545 citations counted in INSPIRE as of 11 Nov 2017

%\cite{Elizalde:2010ep}
\bibitem{Elizalde:2010ep}
E.~Elizalde, S.~Nojiri, S.~D.~Odintsov and D.~Saez-Gomez,
%``Unifying inflation with dark energy in modified F(R) Horava-Lifshitz gravity,''
Eur.\ Phys.\ J.\ C {\bf 70} (2010) 351
doi:10.1140/epjc/s10052-010-1455-7
[arXiv:1006.3387 [hep-th]].
%%CITATION = doi:10.1140/epjc/s10052-010-1455-7;%%
%49 citations counted in INSPIRE as of 11 Nov 2017

%\cite{Elizalde:2010jx}
\bibitem{Elizalde:2010jx}
E.~Elizalde, R.~Myrzakulov, V.~V.~Obukhov and D.~Saez-Gomez,
%``LambdaCDM epoch reconstruction from F(R,G) and modified Gauss-Bonnet gravities,''
Class.\ Quant.\ Grav.\  {\bf 27} (2010) 095007
doi:10.1088/0264-9381/27/9/095007
[arXiv:1001.3636 [gr-qc]].
%%CITATION = doi:10.1088/0264-9381/27/9/095007;%%
%98 citations counted in INSPIRE as of 11 Nov 2017

%\cite{Cognola:2007zu}
\bibitem{Cognola:2007zu}
G.~Cognola, E.~Elizalde, S.~Nojiri, S.~D.~Odintsov, L.~Sebastiani and S.~Zerbini,
%``A Class of viable modified f(R) gravities describing inflation and the onset of accelerated expansion,''
Phys.\ Rev.\ D {\bf 77} (2008) 046009
doi:10.1103/PhysRevD.77.046009
[arXiv:0712.4017 [hep-th]].
%%CITATION = doi:10.1103/PhysRevD.77.046009;%%
%391 citations counted in INSPIRE as of 11 Nov 2017

%\cite{Briscese:2006xu}
\bibitem{Briscese:2006xu}
F.~Briscese, E.~Elizalde, S.~Nojiri and S.~D.~Odintsov,
%``Phantom scalar dark energy as modified gravity: Understanding the origin of the Big Rip singularity,''
Phys.\ Lett.\ B {\bf 646} (2007) 105
doi:10.1016/j.physletb.2007.01.013
[hep-th/0612220].
%%CITATION = doi:10.1016/j.physletb.2007.01.013;%%
%191 citations counted in INSPIRE as of 11 Nov 2017

%\cite{Elizalde:2004mq}
\bibitem{Elizalde:2004mq}
E.~Elizalde, S.~Nojiri and S.~D.~Odintsov,
%``Late-time cosmology in (phantom) scalar-tensor theory: Dark energy and the cosmic speed-up,''
Phys.\ Rev.\ D {\bf 70} (2004) 043539
doi:10.1103/PhysRevD.70.043539
[hep-th/0405034].
%%CITATION = doi:10.1103/PhysRevD.70.043539;%%
%740 citations counted in INSPIRE as of 11 Nov 2017

%\cite{Harko:2011kv}
\bibitem{Harko:2011kv}
T.~Harko, F.~S.~N.~Lobo, S.~Nojiri and S.~D.~Odintsov,
%``$f(R,T)$ gravity,''
Phys.\ Rev.\ D {\bf 84} (2011) 024020
doi:10.1103/PhysRevD.84.024020
[arXiv:1104.2669 [gr-qc]].
%%CITATION = doi:10.1103/PhysRevD.84.024020;%%
%444 citations counted in INSPIRE as of 11 Nov 2017

%\cite{Nojiri:2006gh}
\bibitem{Nojiri:2006gh}
S.~Nojiri and S.~D.~Odintsov,
%``Modified f(R) gravity consistent with realistic cosmology: From matter dominated epoch to dark energy universe,''
Phys.\ Rev.\ D {\bf 74} (2006) 086005
doi:10.1103/PhysRevD.74.086005
[hep-th/0608008].
%%CITATION = doi:10.1103/PhysRevD.74.086005;%%
%556 citations counted in INSPIRE as of 11 Nov 2017

%\cite{Oikonomou:2016jjh}
\bibitem{Oikonomou:2016jjh}
V.~K.~Oikonomou and E.~N.~Saridakis,
%``$f(T)$ gravitational baryogenesis,''
Phys.\ Rev.\ D {\bf 94} (2016) no.12,  124005
doi:10.1103/PhysRevD.94.124005
[arXiv:1607.08561 [gr-qc]].
%%CITATION = doi:10.1103/PhysRevD.94.124005;%%
%15 citations counted in INSPIRE as of 11 Nov 2017

%\cite{Astashenok:2013vza}
\bibitem{Astashenok:2013vza}
A.~V.~Astashenok, S.~Capozziello and S.~D.~Odintsov,
%``Further stable neutron star models from f(R) gravity,''
JCAP {\bf 1312} (2013) 040
doi:10.1088/1475-7516/2013/12/040
[arXiv:1309.1978 [gr-qc]].
%%CITATION = doi:10.1088/1475-7516/2013/12/040;%%
%71 citations counted in INSPIRE as of 11 Nov 2017

%\cite{Astashenok:2017dpo}
\bibitem{Astashenok:2017dpo}
A.~V.~Astashenok, S.~D.~Odintsov and A.~de la Cruz-Dombriz,
%``The realistic models of relativistic stars in $f(R) = R + \alpha R^2$ gravity,''
Class.\ Quant.\ Grav.\  {\bf 34} (2017) no.20,  205008
doi:10.1088/1361-6382/aa8971
[arXiv:1704.08311 [gr-qc]].
%%CITATION = doi:10.1088/1361-6382/aa8971;%%
%8 citations counted in INSPIRE as of 11 Nov 2017

%\cite{Linder:2010py}
\bibitem{Linder:2010py}
E.~V.~Linder,
%``Einstein's Other Gravity and the Acceleration of the Universe,''
Phys.\ Rev.\ D {\bf 81} (2010) 127301
Erratum: [Phys.\ Rev.\ D {\bf 82} (2010) 109902]
doi:10.1103/PhysRevD.81.127301, 10.1103/PhysRevD.82.109902
[arXiv:1005.3039 [astro-ph.CO]].
%%CITATION = doi:10.1103/PhysRevD.81.127301, 10.1103/PhysRevD.82.109902;%%
%433 citations counted in INSPIRE as of 11 Nov 2017

%\cite{Ferraro:2006jd}
\bibitem{Ferraro:2006jd}
R.~Ferraro and F.~Fiorini,
%``Modified teleparallel gravity: Inflation without inflaton,''
Phys.\ Rev.\ D {\bf 75} (2007) 084031
doi:10.1103/PhysRevD.75.084031
[gr-qc/0610067].
%%CITATION = doi:10.1103/PhysRevD.75.084031;%%
%348 citations counted in INSPIRE as of 11 Nov 2017

%\cite{Bamba:2016wjm}
\bibitem{Bamba:2016wjm}
K.~Bamba, S.~D.~Odintsov and E.~N.~Saridakis,
%``Inflationary cosmology in unimodular $F(T)$ gravity,''
Mod.\ Phys.\ Lett.\ A {\bf 32} (2017) no.21,  1750114
doi:10.1142/S0217732317501140
[arXiv:1605.02461 [gr-qc]].
%%CITATION = doi:10.1142/S0217732317501140;%%
%15 citations counted in INSPIRE as of 11 Nov 2017

%\cite{Chen:2010va}
\bibitem{Chen:2010va}
S.~H.~Chen, J.~B.~Dent, S.~Dutta and E.~N.~Saridakis,
%``Cosmological perturbations in f(T) gravity,''
Phys.\ Rev.\ D {\bf 83} (2011) 023508
doi:10.1103/PhysRevD.83.023508
[arXiv:1008.1250 [astro-ph.CO]].
%%CITATION = doi:10.1103/PhysRevD.83.023508;%%
%213 citations counted in INSPIRE as of 11 Nov 2017

%\cite{Myrzakulov:2010tc}
\bibitem{Myrzakulov:2010tc}
R.~Myrzakulov,
%``F(T) gravity and k-essence,''
Gen.\ Rel.\ Grav.\  {\bf 44} (2012) 3059
doi:10.1007/s10714-012-1439-z
[arXiv:1008.4486 [physics.gen-ph]].
%%CITATION = doi:10.1007/s10714-012-1439-z;%%
%76 citations counted in INSPIRE as of 11 Nov 2017

%\cite{Dent:2011zz}
\bibitem{Dent:2011zz}
J.~B.~Dent, S.~Dutta and E.~N.~Saridakis,
%``f(T) gravity mimicking dynamical dark energy. Background and perturbation analysis,''
JCAP {\bf 1101} (2011) 009
doi:10.1088/1475-7516/2011/01/009
[arXiv:1010.2215 [astro-ph.CO]].
%%CITATION = doi:10.1088/1475-7516/2011/01/009;%%
%208 citations counted in INSPIRE as of 11 Nov 2017

%\cite{Cai:2011tc}
\bibitem{Cai:2011tc}
Y.~F.~Cai, S.~H.~Chen, J.~B.~Dent, S.~Dutta and E.~N.~Saridakis,
%``Matter Bounce Cosmology with the f(T) Gravity,''
Class.\ Quant.\ Grav.\  {\bf 28} (2011) 215011
doi:10.1088/0264-9381/28/21/215011
[arXiv:1104.4349 [astro-ph.CO]].
%%CITATION = doi:10.1088/0264-9381/28/21/215011;%%
%247 citations counted in INSPIRE as of 11 Nov 2017

%\cite{Sharif:2011bi}
\bibitem{Sharif:2011bi}
M.~Sharif and S.~Rani,
%``F(T) Models within Bianchi Type I Universe,''
Mod.\ Phys.\ Lett.\ A {\bf 26} (2011) 1657
doi:10.1142/S0217732311036127
[arXiv:1105.6228 [gr-qc]].
%%CITATION = doi:10.1142/S0217732311036127;%%
%84 citations counted in INSPIRE as of 11 Nov 2017

%\cite{Odintsov:2016plw}
\bibitem{Odintsov:2016plw}
S.~D.~Odintsov and V.~K.~Oikonomou,
%``Singular $F(R)$ Cosmology Unifying Early and Late-time Acceleration with Matter and Radiation Domination Era,''
Class.\ Quant.\ Grav.\  {\bf 33} (2016) no.12,  125029
doi:10.1088/0264-9381/33/12/125029
[arXiv:1602.03309 [gr-qc]].
%%CITATION = doi:10.1088/0264-9381/33/12/125029;%%
%11 citations counted in INSPIRE as of 11 Nov 2017

%\cite{Odintsov:2015gba}
\bibitem{Odintsov:2015gba}
S.~D.~Odintsov and V.~K.~Oikonomou,
%``Singular Inflationary Universe from $F(R)$ Gravity,''
Phys.\ Rev.\ D {\bf 92} (2015) no.12,  124024
doi:10.1103/PhysRevD.92.124024
[arXiv:1510.04333 [gr-qc]].
%%CITATION = doi:10.1103/PhysRevD.92.124024;%%
%34 citations counted in INSPIRE as of 11 Nov 2017

%\cite{Odintsov:2015zza}
\bibitem{Odintsov:2015zza}
S.~D.~Odintsov and V.~K.~Oikonomou,
%``Bouncing cosmology with future singularity from modified gravity,''
Phys.\ Rev.\ D {\bf 92} (2015) no.2,  024016
doi:10.1103/PhysRevD.92.024016
[arXiv:1504.06866 [gr-qc]].
%%CITATION = doi:10.1103/PhysRevD.92.024016;%%
%45 citations counted in INSPIRE as of 11 Nov 2017

%\cite{Bamba:2008ut}
\bibitem{Bamba:2008ut}
K.~Bamba, S.~Nojiri and S.~D.~Odintsov,
%``The Universe future in modified gravity theories: Approaching the finite-time future singularity,''
JCAP {\bf 0810} (2008) 045
doi:10.1088/1475-7516/2008/10/045
[arXiv:0807.2575 [hep-th]].
%%CITATION = doi:10.1088/1475-7516/2008/10/045;%%
%275 citations counted in INSPIRE as of 11 Nov 2017

%\cite{Beltran:2016jb}
\bibitem{Beltran:2016jb}
Jose Beltran Jimenez, Diego Rubiera-Garcia, Diego Saez-Gomez, and Vincenzo Salzano,
%``Cosmological future singularities in interacting dark energy models,''
Phys.\ Rev.\ D {\bf 94} (2016) no.12, 123520
doi:10.1103/PhysRevD.94.123520
[arXiv:1607.06389 [gr-qc]].
%%CITATION = doi:10.1103/PhysRevD.94.123520;%%
%10 citations counted in INSPIRE as of 19 March 2018

%\cite{Khurshudyan:2014yoa}
\bibitem{Khurshudyan:2014yoa}
M.~Khurshudyan, B.~Pourhassan, R.~Myrzakulov and S.~Chattopadhyay,
%``An effective quintessence field with a power-law potential,''
Astrophys.\ Space Sci.\  {\bf 356} (2015) no.2,  383
doi:10.1007/s10509-014-2209-z
[arXiv:1403.3768 [gr-qc]].
%%CITATION = doi:10.1007/s10509-014-2209-z;%%
%10 citations counted in INSPIRE as of 11 Nov 2017

%\cite{Zhang:2016zh}
\bibitem{Zhang:2016zh}
Ming-Jian Zhang, Jun-Qing Xia,
%``Test of the cosmic evolution using Gaussian processes,''
JCAP {\bf 1612} (2016) no.12, 005
doi:10.1088/1475-7516/2016/12/005
[arXiv:1606.04398 [astro-ph.CO]].
%%CITATION = doi:10.1088/1475-7516/2016/12/005;%%
%6 citations counted in INSPIRE as of 19 March 2018

%\cite{Yahya:2014s}
\bibitem{Yahya:2014s}
S.~Yahya, M.~Seikel, C.~Clarkson, R.~Maartens and M.~Smith,
%``Null tests of the cosmological constant using supernovae,''
Phys.Rev. D89 (2014) no.2, 023503
doi:10.1103/PhysRevD.89.023503
[arXiv:1308.4099 [astro-ph.CO] ].
%%CITATION = doi:10.1103/PhysRevD.89.023503 ;%%
%25 citations counted in INSPIRE as of 19 March 2018


%\cite{Santos:2015s}
\bibitem{Santos:2015s}
S.~Santos-da Costa, V.~C.~Busti and R.~F.~L.~Holanda,
%``Two new tests to the distance duality relation with galaxy clusters,''
JCAP 1510 (2015) no.10, 061
doi:10.1088/1475-7516/2015/10/061
[arXiv:1506.00145 [astro-ph.CO] ].
%%CITATION = doi:10.1088/1475-7516/2015/10/061;%%
%15 citations counted in INSPIRE as of 19 March 2018

%\cite{Holsclaw:2010th}
\bibitem{Holsclaw:2010th}
T.~Holsclaw, U.~Alam, B.~Sanso, H.~Lee, K.~Heitmann, S.~Habib, D.~ Higdon,
%``Nonparametric dark energy reconstruction from supernova data,''
Phys. Rev. Lett. {\bf 105} (2010) 241302
doi:10.1103/PhysRevLett.105.241302
[arXiv:1011.3079 [astro-ph.CO] ].
%%CITATION = doi:10.1103/PhysRevLett.105.241302;%%
%81 citations counted in INSPIRE as of 19 March 2018

%\cite{Seikel:2012mcm}
\bibitem{Seikel:2012mcm}
M.~Seikel, C.~Clarkson and M.~Smith,
%``Reconstruction of dark energy and expansion dynamics using Gaussian processes,''
JCAP {\bf 1206} (2012) 036
doi:10.1088/1475-7516/2012/06/036
[arXiv:1204.2832 [astro-ph.CO] ].
%%CITATION = doi:10.1088/1475-7516/2012/06/036;%%
%86 citations counted in INSPIRE as of 19 March 2018

%\cite{Shafieloo:2012as}
\bibitem{Shafieloo:2012as}
A.~Shafieloo, A.~G.~Kim and E.~V.~Linder,
%``Gaussian process cosmography,''
Phys.\ Rev.\ D {\bf 85} (2012) 123530
doi:10.1103/PhysRevD.85.123530
[arXiv:1204.2272 [astro-ph.CO] ].
%%CITATION = doi:10.1103/PhysRevD.85.123530;%%
%45 citations counted in INSPIRE as of 19 March 2018

%\cite{Yang:2015tzr}
\bibitem{Yang:2015tzr}
T.~Yang, Z.~K.~Guo and R.~G.~Cai,
%``Reconstructing the interaction between dark energy and dark matter using Gaussian processes,''
Phys.\ Rev.\ D {\bf 91} (2015) no.12, 123533
doi:10.1103/PhysRevD.91.123533
[arXiv:1505.04443 [astro-ph.CO] ].
%%CITATION = doi:110.1103/PhysRevD.91.123533;%%
%33 citations counted in INSPIRE as of 19 March 2018

%\cite{Dabrowski:2009kg}
\bibitem{Dabrowski:2009kg}
M.~P.~Dabrowski and T.~Denkiewicz,
%``Barotropic index w-singularities in cosmology,''
Phys.\ Rev.\ D {\bf 79} (2009) 063521
doi:10.1103/PhysRevD.79.063521
[arXiv:0902.3107 [gr-qc]].
%%CITATION = doi:10.1103/PhysRevD.79.063521;%%
%65 citations counted in INSPIRE as of 11 Nov 2017

%\cite{FernandezJambrina:2007sx}
\bibitem{FernandezJambrina:2007sx}
L.~Fernandez-Jambrina,
%``Hidden past of dark energy cosmological models,''
Phys.\ Lett.\ B {\bf 656} (2007) 9
doi:10.1016/j.physletb.2007.08.091
[arXiv:0704.3936 [gr-qc]].
%%CITATION = doi:10.1016/j.physletb.2007.08.091;%%
%27 citations counted in INSPIRE as of 11 Nov 2017

%\cite{Zheng:2016jlq}
\bibitem{Zheng:2016jlq}
X.~Zheng, X.~Ding, M.~Biesiada, S.~Cao and Z.~Zhu,
%``What are $Omh^2(z_1,z_2)$ and $Om(z_1,z_2)$ diagnostics telling us in light of $H(z)$ data?,''
Astrophys.\ J.\  {\bf 825} (2016) no.1,  17
doi:10.3847/0004-637X/825/1/17
[arXiv:1604.07910 [astro-ph.CO]].
%%CITATION = doi:10.3847/0004-637X/825/1/17;%%
%13 citations counted in INSPIRE as of 11 Nov 2017

%\cite{VarunSahni:2014}
\bibitem{VarunSahni:2014}
V. Sahni, A. Shafieloo, A. A. Starobinsky,
%``Model independent evidence for dark energy evolution from Baryon Acoustic Oscillations,''
ApJL {\bf 793} (2014) no.2, L40
doi:10.1088/2041-8205/793/2/L40
[arXiv:1406.2209 [astro-ph.CO]].
%%CITATION = https://doi.org/10.1088/2041-8205/793/2/L40
%%83 citations counted in INSPIRE as of 12 Nov 2017

%\cite{VarunSahni:2008}
\bibitem{VarunSahni:2008}
Varun Sahni, Arman Shafieloo, Alexei A. Starobinsky,
%``Two new diagnostics of dark energy,''
Phys.\ Rev.\ D {\bf 78} (2008) 103502
doi:10.1103/PhysRevD.78.103502
[arXiv:0807.3548 [astro-ph]].
%%CITATION = https://doi.org/10.1103/PhysRevD.78.103502;%%
%277 citations counted in INSPIRE as of 12 Nov 2017



\end{thebibliography}
\end{document}